\documentclass[a4paper,11pt]{article}
\usepackage{jheppub}
\usepackage[T1]{fontenc}
\usepackage{lmodern,booktabs}
\hypersetup{pdftitle={Microcausality Bounds on Dispersion Relations},
            pdfauthor={Enrico Pajer and Borna Salehian}}
\title{Microcausality Bounds on Real Dispersion Relations}
\author[a]{Enrico Pajer}
\author[b]{and Borna Salehian}
\affiliation[a]{Department of Applied Mathematics and Theoretical Physics,
University of Cambridge,\\
Wilberforce Road, Cambridge, CB3 0WA, UK}
\affiliation[b]{Institut f\"ur Theoretische Physik, ETH Z\"urich,\\
Wolfgang-Pauli-Strasse 27, 8093 Z\"urich, Switzerland}
\emailAdd{enrico.pajer@gmail.com,bsalehian@phys.ethz.ch}
\abstract{
We constrain non-relativistic dispersion relations from relativistic causality. For a non-dissipative dispersion branch $\omega(k)$ that is analytic in a disk of complex momentum, microcausality requires $|\Im\omega(k)|\le|\Im k|$. We show that such branches are in one-to-one correspondence with real functions bounded between $-1$ and
$1$, which we call profiles. Taylor coefficients of the branch expansion inside a complex-momentum convergence disk are moments of the profile, turning causal constraint into a Markov \(L\)-moment problem. We derive sharp analytic bounds on each coefficient, selected joint bounds on pairs and triplets and a simple sphere bound on the sum of squares. Saturating any of these bounds fixes the entire branch, but violates microcausality outside the disk. Measuring one or more coefficients places an upper bound of the radius of convergence of the dispersion relation. 
}

\usepackage{physics}
\usepackage{tcolorbox}
\definecolor{greyish2}{rgb}{.96,.96,.96}

\begin{document}
\maketitle

\section{Introduction}
\label{sec:introduction}

Causality requires a system to respond to a perturbation only after it
has been applied. In Fourier space this is the origin of the analyticity of response functions in the upper half of the complex frequency plane, and of the Kramers--Kronig relations. Relativity strengthens this requirement: the response
must remain inside the future light cone of the perturbation. In quantum field theory, this follows from microcausality, namely the
vanishing of commutators of local operators at spacelike separation, and it enlarges the domain of analyticity from complex frequency to complex frequency \emph{and} momentum
\cite{Leontovich1961,Dolgov:1981,Streater:1989vi,Nussenzveig:1972tcd,bogolubov2012general}. In this paper we study what this domain of analyticity implies for the dispersion relations of collective excitations and for the coefficients of their low-momentum expansion.

Microcausality is an operator statement and therefore holds in every
state of the theory (see \cite{Dubovsky:2007ac,Hartman:2015lfa,Hui:2025aja,Creminelli:2025rxj}). In particular, it continues to constrain states
that break Lorentz boosts, such as those at finite temperature or
density, whose excitations can have dispersion relations
quite different from those of relativistic particles in vacuum. Nevertheless, these
excitations appear as singularities of a retarded Green function $G_{R}$, whose analyticity domain is given by microcausality. Let $ \omega(k)$ be a dispersion branch describing a pole of $ G_{R}(\omega,\mathbf{k})$, where $k$ denotes momentum along a fixed spatial direction. Restricting to real $k$, one might guess that causality is the statement that the group velocity is subluminal. This is not a valid criterion in general: causal dissipative systems can have superluminal group velocity, and subluminal group velocity does not guarantee causality \cite{Brillouin1960,Fox:1970cu,Krotscheck:1978,Pu:2009fj,Gavassino:2023mad,Hoult:2023clg,Wang:2023csj}. For non-dissipative branches, which are real for real $k$, a subluminal group velocity is indeed necessary \cite{Bros:2001ab,Hui:2025aja}, a harbinger of the more rigid structure for real dispersion branches. However a subluminal group velocity is far from sufficient. Constraints from complex momentum are much stronger, and there the dispersion relation behaves very differently from its restriction to the real axis. 

\begin{figure}[t]
 \centering
 \includegraphics[width=\textwidth]{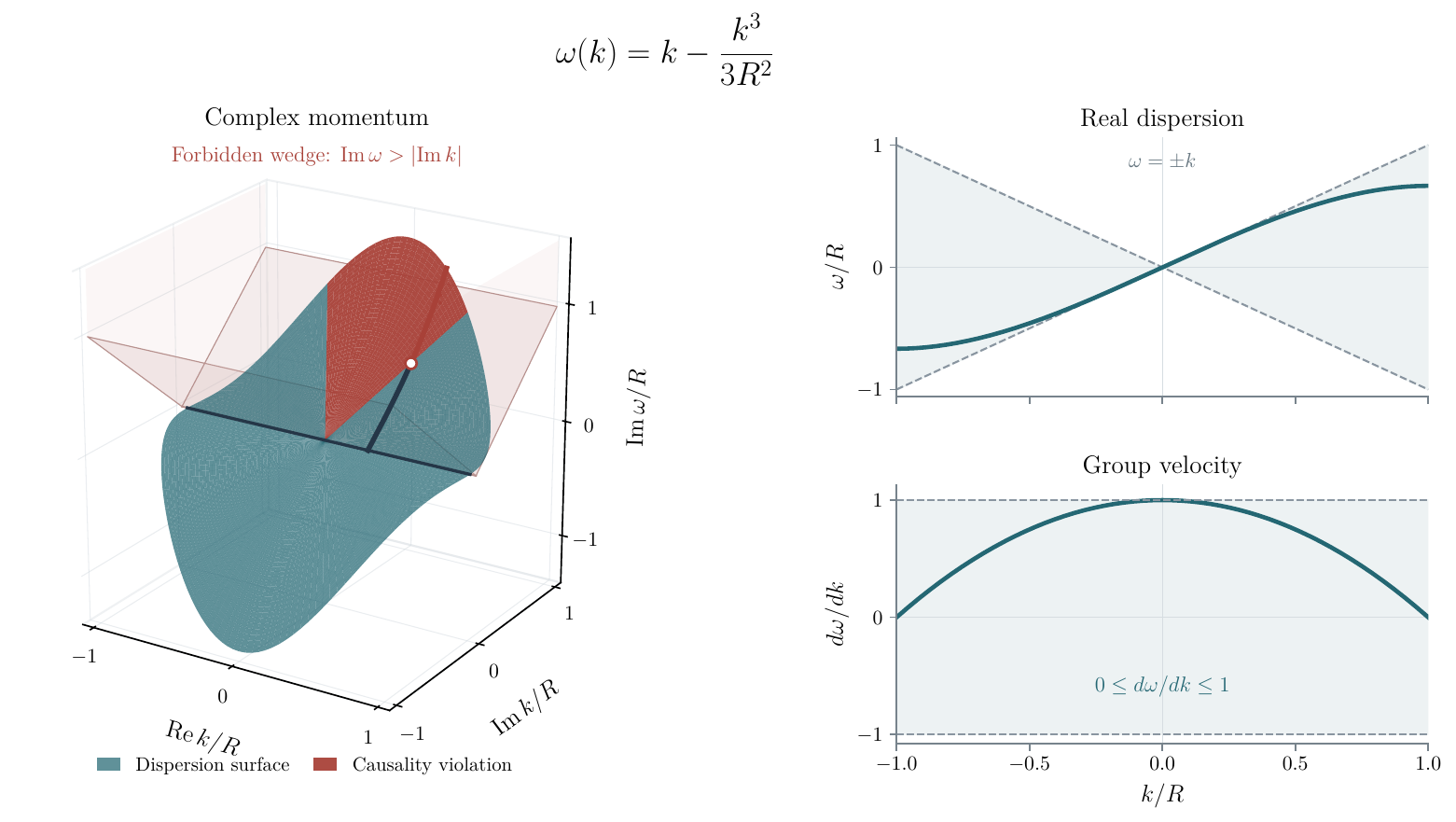}
 \caption{Subluminal group velocity does not ensure microcausality.
 The cubic $\omega(k)=k-k^3/(3R^2)$ satisfies
 $|\omega(k)|\leq|k|$ and $0\leq d\omega/dk\leq1$ for real $|k|\leq R$,
 as shown on the right. Its complex continuation nevertheless enters
 the forbidden wedge $\operatorname{Im}\omega>|\operatorname{Im}k|$.
 The left panel shows $\operatorname{Im}\omega(k)$ over the disk
 $|k|\leq R$, with the violating part of the surface in red.
 The thick curve follows the surface at fixed
 $\operatorname{Re}k=3R/10$ as $\operatorname{Im}k$ increases.
 It changes from blue to red at the white marker, where
 $\operatorname{Im}k=3\sqrt3 R/10$ and
 $\operatorname{Im}\omega=\operatorname{Im}k$.
 Here $R$ specifies the displayed momentum scale; the polynomial
 itself has an infinite Taylor radius.}
 \label{fig:paper-forbidden-cubic}
\end{figure}

Indeed, for stationary, homogeneous states, the Green function is analytic in the region $\operatorname{Im}\omega>|\operatorname{Im}k|$ and therefore the dispersion relation must satisfy 
\begin{equation}
 \operatorname{Im}\omega(k)\leq|\operatorname{Im}k|\,.
 \label{eq:introduction-causality}
\end{equation}
This inequality provides a direct route from relativistic causality to
constraints on non-relativistic dispersion relations. Figure~\ref{fig:paper-forbidden-cubic} illustrates the improvement obtained by considering the complex plane, focusing on a simple non-dissipative $ \omega(k)$ that violates microcausality. Throughout the displayed real-momentum interval, the cubic dispersion lies between $\omega=\pm k$ and has group velocity between
zero and one. Its continuation nevertheless enters the forbidden region
$\operatorname{Im}\omega>|\operatorname{Im}k|$ within the corresponding
complex disk. 

The consequences of relativistic causality for hydrodynamics have been
studied extensively. It includes the acausality and instability of first-order theories and their resolutions
\cite{Israel:1979wp,Hiscock:1983zz,Hiscock:1985zz,Bemfica:2017wps,Kovtun:2019hdm,Bemfica:2019knx,Hoult:2020eho,Bemfica:2020zjp}, the relation between causality and stability
\cite{Gavassino:2021owo,Gavassino:2023myj}, and bounds on transport from the existence of a light cone \cite{Hartman:2017hhp}. The implications of
microcausality for Green functions and effective field theories on
Lorentz-breaking states have been worked out in
\cite{Baumann:2015nta,Grall:2021xxm,Creminelli:2022onn,Hui:2023pxc,Creminelli:2023kze,Creminelli:2024lhd,Hui:2025aja,Creminelli:2025rxj,Cheung:2026wrz}. More recently, Heller, Serantes, Spali\'nski and Withers developed a systematic approach
based on complex-momentum analyticity, deriving bounds on transport
coefficients and on the analytic structure of dispersion
relations~\cite{Heller:2022ejw,Heller:2023hydrohedron}. Most recently, Creminelli, Mancini and McBlain improved the bounds of
\cite{Heller:2022ejw,Heller:2023hydrohedron} for non-dissipative
dispersion relations using the theory of typically real functions (see e.g. \cite{Rogosinski1932,Robertson1935}),
obtaining coefficient bounds and an infinite hierarchy of determinant
constraints~\cite{Creminelli:2026dispersion}. Their formulation provides
the starting point for this work, which began after learning about their interesting results. Investigating the sharpness of their bounds led us to a complementary description in terms of moments of a bounded function as outlined below. 

\begin{figure}
\centering
\includegraphics[width=\linewidth]{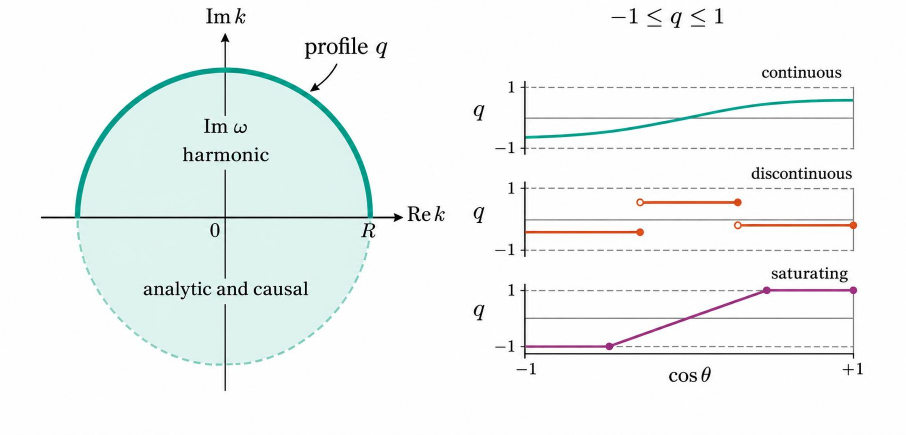}
\caption{A profile \(q=\operatorname{Im}\omega/\operatorname{Im}k\),
bounded by \(|q|\leq 1\) on the upper semicircle, determines an analytic dispersion branch causal within \(|k|<R\).
The right panels illustrate continuous and discontinuous profiles,
and a profile attaining \(q=\pm 1\) on intervals.}
    \label{fig:intro}
\end{figure}
\paragraph{Summary of the main results.}
For the convenience of the reader, we summarize below the main results
of our paper. We write a real analytic dispersion branch as\footnote{A possible real gap $ \omega(0)=\omega_{0}$ of the dispersion relation drops out from all bounds we derive, and hence we omit it throughout. Our bounds are independent of this choice.}
\begin{equation}
 \omega(k)=R\sum_{n=1}^{\infty}a_n(k/R)^n\,,
 \qquad a_1=c_s\,,\qquad a_n=\alpha_nR^{n-1}\quad(n\geq2)\,.
 \label{eq:introduction-normalized-series}
\end{equation}
Assuming that this branch remains on the physical sheet for all values of $k$ in the disk $|k|<R$, we impose
\eqref{eq:introduction-causality} and derive the following bounds.
\begin{enumerate}
 \item \emph{Moments of a bounded function.}
 At fixed $R$, we establish a map between causal dispersion relations and a bounded function $-1\leq q(x)\leq1$ for $-1\leq x\leq 1$. We call this function the causal profile or simply the \textit{profile} and it is the workhorse of our construction. It is defined as
 \[ q(\cos \theta):=\left[\frac{\operatorname{Im} \omega(k)}{\operatorname{Im}k}\right]_{k=R e^{i\theta}}\,, \]
 and depicted in Fig.~\ref{fig:intro}. The Taylor coefficients $ a_{n}$ are moments of $q$, and the explicit integral in \eqref{eq:paper-profile-reconstruction} reconstructs the branch from the profile. This is a bounded moment problem, also called the Markov
 $L$-problem~\cite{KreinNudelman:1977}. Section~\ref{sec:moment-problem} gives the coefficient-extraction and
 reconstruction formulas and proves causality of the reconstructed branch.

 \item \emph{Individual coefficients and extremal dispersion relations.}
 The sharp bounds are
 \begin{equation}
  |c_s|\leq1\,,\qquad
  |a_n|\leq M_n=\frac{4n}{\pi(n^2-1)}\cot\frac{\pi}{2n}\,,
  \quad (n\geq2)\,.
  \label{eq:introduction-individual-bounds}
 \end{equation}
 The bound on $c_s$ and the absence of higher-order corrections at
 $c_s=\pm1$ were already established in Ref.~\cite{Heller:2022ejw}
 and discussed in Ref.~\cite{Hui:2025aja,Creminelli:2026dispersion}. At every order,
 maximizing or minimizing one coefficient uniquely fixes the profile
 and hence all the other Taylor coefficients. We give explicit
 all-order extremal dispersion relations in
 Section~\ref{sec:paper-individual-bounds}.

 \item \emph{A complete family of linear bounds.}
 The allowed region for the Taylor coefficients is compact and convex. We derive
 sharp bounds on arbitrary finite linear combinations of the
 coefficients and evaluate them in terms of elementary functions at polynomial roots. The full family of linear bounds is necessary and
 sufficient for a set of Taylor coefficient to admit a causal analytic
 branch on the disk (see Section~\ref{sec:paper-linear-combinations}).

 \item \emph{The sphere bound.}
 Parseval's theorem gives the sharp bound
 \begin{equation}
  \sum_{n=1}^{\infty}a_n^2\leq1\,.
  \label{eq:introduction-sphere-bound}
 \end{equation}
 This is necessary but not sufficient. If the first $N$ coefficients vanish, the remaining bound
 $\sum_{n=N+1}^{\infty}a_n^2\leq1$ is still sharp for every finite $N$.
 Equality holds precisely for profiles taking only the values minus one and one almost everywhere. 

 \item \emph{Exact pair and triple regions.}
 We obtain analytic parametric bounds for $(c_s,a_n)$ at every $n$.
 For odd coefficients, the sharp regions were already characterized
 using typically real functions and computed numerically by
 Creminelli, Mancini and McBlain~\cite{Creminelli:2026dispersion}.
 Our expressions give analytic boundaries agreeing with their
 converged numerical regions for $n=3,5,7,9$ in their Figure~6.
 We also treat even coefficients, and obtain necessary and
 sufficient nonlinear bounds for $(c_s,a_2,a_3)$ (see
 Sections~\ref{sec:paper-parametric-pairs}--\ref{sec:paper-three-coefficients}).

 \item \emph{The convergence radius.}
 The map between bounded profiles and dispersion relations guarantees $R_{\mathrm{conv}}\geq R$, with equality
 for branches extremizing a coefficient of order $n\geq2$ and for non-linear branches saturating the sphere bound.
 Conversely, if the branch obeys causality throughout its true Taylor
 disk, each known nonzero coefficient gives
 \begin{equation}
  R_{\mathrm{conv}}\leq
  \left(\frac{M_n}{|\alpha_n|}\right)^{1/(n-1)},\qquad n\geq2.
  \label{eq:introduction-radius-bound}
 \end{equation}
 Several known coefficients give the minimum of their upper bounds;
 knowing all orders recovers the true radius. These results are
 discussed in Section~\ref{sec:convergence-radius}.
 \item \emph{Strict yet sharp bounds.} Dispersion relations can be continued analytically outside the convergence disk. If the profile $q(x)$ equals \(+1\) or \(-1\) in an open interval, this continuation violates microcausality immediately outside the disc, with the exception for the linear branches \(\omega(k)=\pm k\). The profiles extremizing all of our bounds have this property and are hence not globally causal. This turns our bounds into strict bounds, for example \(|a_n|<M_n\) for every \(n\ge2\). The strict bounds remain sharp: for each fixed \(n\), we construct globally-causal rational
Green functions whose Taylor coefficients approach these bounds arbitrarily closely (see Section~\ref{sec:global-causality}).
\end{enumerate}

\section{Retarded Green function and the dispersion branch}
\label{sec:retarded-dispersion}

A state that breaks Lorentz boosts can support excitations with dispersion
relations quite different from those of particles in vacuum. The state
may select a preferred rest frame, a speed of sound, or a scale at which
dispersion becomes important. Nevertheless, the underlying light cone
still limits how a disturbance can propagate. To extract this restriction
without specifying the microscopic dynamics, we begin with the retarded
response and ask where its poles can lie.

\subsection{Retarded response and microcausality}
\label{sec:retarded-response}

We work in Minkowski spacetime and consider a stationary, homogeneous
state.\footnote{Time translation may be spontaneously broken, provided that it is
broken together with an internal symmetry into a diagonal subgroup that
plays the same role. The standard example is a system at
finite chemical potential $\mu$ for a $U(1)$ charge $Q$, such as a superfluid \cite{Son:2002zn,Nicolis:2011pv}.} This state may be a pure state or a mixed state.
For a local Hermitian bosonic operator $\mathcal O$, define
\begin{equation}
 G_R(t,\mathbf x)
 =i\Theta(t)\langle[\mathcal O(t,\mathbf x),\mathcal O(0,\mathbf0)]\rangle\,.
 \label{eq:retarded-definition}
\end{equation}
Here $\Theta$ is the Heaviside step function. The expectation value is
taken in the chosen state, which need not be
invariant under boosts or rotation. We set the microscopic light-cone speed to one. Microcausality states that local operators commute at spacelike separation; this gives
\begin{equation}
 G_R(t,\mathbf x)=0\qquad\text{for}\qquad |\mathbf x|>t\,.
 \label{eq:causal-support}
\end{equation}
A disturbance at the origin therefore cannot affect a point at distance
$|\mathbf x|$ before time $t=|\mathbf x|$. We will express this condition
in frequency and momentum space, where it constrains the locations of
poles and hence the dispersion relations.

We use the Fourier convention
\begin{equation}
 G_R(\omega,\mathbf k)=\int \dd{t}\,\dd[3]{\mathbf x}\,
       e^{i\omega t-i\mathbf k\cdot\mathbf x}G_R(t,\mathbf x)\,,
 \label{eq:retarded-fourier}
\end{equation}
and assume that the position-space response is a tempered distribution. This means that $G_R(t,\mathbf x)$ is polynomially bounded and the Fourier transform, for real $\omega$ and $\mathbf k$, is also a tempered distribution (See \cite{Creminelli:2025rxj}). In particular, an exponentially growing response is excluded. We now allow $\omega$ and $\mathbf k$ to be complex. Since
\eqref{eq:causal-support} restricts the nonzero response to
$t\geq|\mathbf x|$, the Fourier kernel obeys
\begin{equation}
 \left|e^{i\omega t-i\mathbf k\cdot\mathbf x}\right|
 =e^{-\operatorname{Im}\omega\,t+\operatorname{Im}\mathbf k\cdot\mathbf x}
 \leq e^{-(\operatorname{Im}\omega-|\operatorname{Im}\mathbf k|)t}\,.
 \label{eq:fourier-damping}
\end{equation}
The exponential suppresses contributions from large separations whenever
$\operatorname{Im}\omega>|\operatorname{Im}\mathbf k|$. Consequently,
the Fourier transform is analytic when
$(\operatorname{Im}\omega,\operatorname{Im}\mathbf k)$ lies inside the
future light cone. We call the corresponding domain of complex frequency
and momentum, with arbitrary real parts, the \emph{future tube}:\footnote{We preferred this terminology over forward lightcone, used for instance in \cite{Creminelli:2022onn,Creminelli:2024lhd}, to keep the distinction between this and position space forward lightcone.}
\begin{equation}
 G_R(\omega,\mathbf k)\ \text{is analytic for}\quad
 \operatorname{Im}\omega>|\operatorname{Im}\mathbf k|\,.
 \label{eq:forward-tube}
\end{equation}
In the case of many fields, $G_R$ becomes a matrix of retarded
propagators, and the same analyticity condition applies to every matrix
element. However, we will not study that possibility here.
We follow the use of this analyticity property to constrain dispersion
relations developed in Refs.~\cite{Heller:2022ejw,Heller:2023hydrohedron,Hui:2025aja,Creminelli:2026dispersion}.

\subsection{Analytic dispersion branches}
\label{sec:analytic-branches}

Fixing a spatial direction, say $\mathbf k=(k,0,0)$, gives a signed momentum
component $k\in\mathbb R$ that we can continue to one complex variable.
This choice keeps branches such as
$\omega=\pm c_s k$ analytic at the origin, whereas using $k^2$ as the
variable would not. Note that isotropy is not required for our bounds along a fixed momentum direction, e.g.~the $ x  $ direction here. For an isotropic state this is all
there is, since the dispersion relation is the same in every direction. However, in an anisotropic state, these bounds apply separately in each direction, with direction-dependent coefficients and analyticity radii.

A dispersion branch $\omega(k)$ specifies a pole of the single retarded
Green function $G_R(\omega,k)$, evaluated along this momentum line. It solves
\begin{equation}
 G_R^{-1}(\omega(k),k)=0\,.
 \label{eq:inverse-zero}
\end{equation}
We consider poles on the physical sheet of the retarded
Green function, where \eqref{eq:forward-tube} applies. This is an essential assumption. If $G_R$ has
branch cuts, a pole can pass through a cut onto another sheet as $k$ is continued into the complex plane. The function $\omega(k)$ may remain analytic there, but it is then a pole of the analytic continuation of $G_R$ to another sheet,
and no longer describes a singularity of $G_R$ itself. It can therefore enter the
future tube without contradicting \eqref{eq:forward-tube}, and the inequality below
need not hold\footnote{Kinetic theory, where the non-hydrodynamic sector is a branch cut, provides explicit examples \cite{Heller:2020hnq}. Another example is the colour-superconductor studied in \cite{Creminelli:2026dispersion}.} \cite{Gavassino:2026acausal,Brants:2026causalUV,Creminelli:2026dispersion}. Throughout, we will restrict to a domain within which the branch remains on the physical sheet.

The absence of poles in the future tube immediately implies
\begin{equation}
 \operatorname{Im}\omega(k)\leq|\operatorname{Im}k|\,.
 \label{eq:one-sided-causality}
\end{equation}
This is a necessary consequence of microcausality. By itself it constrains neither the residues, nor the other singularities of $G_R$, nor its behaviour at large complex $(\omega,\mathbf k)$.\footnote{Analyticity in the future tube is necessary but not sufficient for \eqref{eq:causal-support}: a function analytic in the future tube is the Fourier transform of a tempered distribution supported in the future light cone if and only if it is also polynomially bounded there, both at large $|\omega|,|\mathbf k|$ and as the boundary of the tube is approached \cite{Vladimirov1966,bogolubov2012general,Creminelli:2025rxj}.} These additional questions will matter when we discuss
realization in Section~\ref{sec:global-causality}. 

To turn \eqref{eq:one-sided-causality} into constraints on coefficients,
we assume that the selected branch is analytic around $k=0$ and write
\begin{equation}
 \omega(k)=c_s k+\sum_{n=2}^{\infty}\alpha_n k^n,
 \quad\text{for}\quad |k|<R\,.
 \label{eq:paper-dispersion-series}
\end{equation}
At this stage the coefficients may be complex. The notation $c_s$ denotes
the linear coefficient even when the mode is not sound. Here $R>0$ is the radius of a disk on which both analyticity and microcausality, i.e.
\eqref{eq:one-sided-causality}, hold; it need not initially be identified with the actual radius of convergence of the Taylor series.

It is worth stressing that analyticity of the branch is a separate assumption from analyticity of the
Green function away from its poles. If the inverse Green function is analytic
near $(\omega_0,0)$ and
$\partial_\omega G_R^{-1}(\omega_0,0)\ne0$, the implicit function theorem
guarantees an analytic branch. At a multiple zero one must check this property directly. See also \cite{Grozdanov:2019uhi}. For example, for the equation $\omega^2-c_s^2k^2=0$ we can label branches in a useful and less useful way. For the former, we can take two roots $\omega=\pm c_s k$. For each of them $  \omega $ is analytic in $  k $, despite the fact that the two branches cross at the origin. For the latter, for positive real $c_s$, the positive-frequency selection on the
real axis is instead $c_s|k|$. This is not a useful choice for us because it is not an analytic branch at $  k=0 $. When no such labeling exists, as for $\omega^2\propto k^3$, our analysis does not apply. 


\subsection{Dispersion relations and their orbits}
\label{sec:paper-branch-symmetries}

Here we discuss when a branch is related to other branches by the reality of the retarded Green function or other symmetries. Since the commutator of two Hermitian operators has a purely imaginary
expectation value,
$G_R(t,\mathbf x)$ is real, which in turn relates its pole branches. In Fourier space it satisfies
\begin{equation}
 [G_R(\omega,k)]^*=G_R(-\omega^*,-k^*)\,.
 \label{eq:hermitian-reality}
\end{equation}
A pole at $(\omega,k)$ is
therefore accompanied by a pole at $(-\omega^*,-k^*)$. For an analytic
dispersion branch, the corresponding map is
\begin{equation}
 \omega(k)\ \longmapsto\ -[\omega(-k^*)]^*\,.
 \label{eq:paper-reality-partner}
\end{equation}
Applying it twice returns the original branch. The orbit, meaning the
set of distinct branches obtained by applying the transformation, thus
contains one or two members. Note that reality alone does not necessarily imply two distinct branches: the transformed function may coincide with the original one, such as for example $ \omega=c_{s}k+\alpha_{3}k^{3}$ with $ c_{s}$ and $ \alpha_{3} $ real.

The theory and state may have additional symmetries and these may relate a given branch to other new branches. For example, consider inversion along the chosen momentum direction,
$G_R(\omega,k)=G_R(\omega,-k)$. This can follow from isotropy or parity,
provided the symmetry leaves the chosen correlator invariant. We denote
this inversion by $P$ and the reality transformation
\eqref{eq:paper-reality-partner} by $CP$. Their composition defines $C$,
giving the maps
\begin{equation}
 (P\omega)(k)=\omega(-k)\,,\quad
 (C\omega)(k)=-[\omega(k^*)]^*\,,\quad
 (CP\omega)(k)=-[\omega(-k^*)]^*\,.
 \label{eq:paper-branch-maps}
\end{equation}

\begin{figure}[!t]
 \centering
 \includegraphics[width=\textwidth]{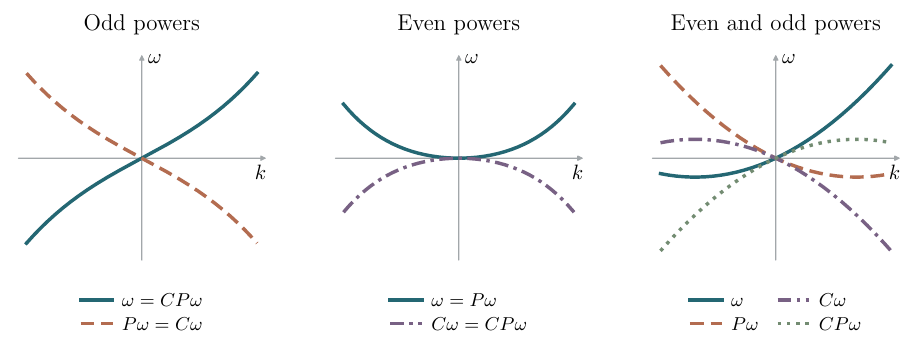}
 \caption{Odd, even, and mixed real dispersion branches have respectively
 two, two, and four images under $C$ and $P$. The examples obey the causal inequality on a common disk.}
 \label{fig:paper-branch-orbits}
\end{figure}

Note that they preserve analyticity of  a given dispersion relation and obey $P^2=C^2=1$ and $PC=CP$.
The resulting group has four elements, $\{1,P,C,CP\}$, i.e. it is
$\mathbb Z_2\times\mathbb Z_2$. The map $C$ is defined here through
reality and inversion; identifying it with a physical charge-conjugation
symmetry is not required. The orbit of a branch is
\begin{equation}
 \bigl\{\omega(k),\ \omega(-k),\ -[\omega(k^*)]^*,
                  \ -[\omega(-k^*)]^*\bigr\}\,.
 \label{eq:paper-complex-orbit}
\end{equation}
It can contain one, two, or four distinct branches, since the orbit size
must divide the group order. Figure~\ref{fig:paper-branch-orbits} shows
examples with real coefficients and two or four distinct branches.
For unrestricted Taylor coefficients the minimum number of branches remains one: for example $\omega(k)=\sum_{n}\alpha_{2n}k^{2n}$, with purely imaginary $\alpha_{2n}$, is fixed by all four maps. Instead, if we restrict ourselves to only real Taylor coefficients, as we will do starting from the next subsection and for the rest of the paper, then the minimum number of branches is two. We also stress that the full Green function may contain several other branches, each with their corresponding orbits.

A useful result is that it is sufficient to impose causality on one dispersion relation within an orbit; all the others then automatically satisfy it. Suppose \eqref{eq:one-sided-causality} holds for
$\omega(k)$ throughout $|k|<R$. The maps $k\mapsto-k$, $k\mapsto k^*$,
and $k\mapsto-k^*$ preserve this disk. Using
$\operatorname{Im}(-z^*)=\operatorname{Im}z$, we find
\begin{equation}
 \begin{aligned}
 \operatorname{Im}(P\omega)(k)
 &=\operatorname{Im}\omega(-k)
 \leq|\operatorname{Im}(-k)|=|\operatorname{Im}k|\,,\\
 \operatorname{Im}(C\omega)(k)
 &=\operatorname{Im}\omega(k^*)
 \leq|\operatorname{Im}k^*|=|\operatorname{Im}k|\,,\\
 \operatorname{Im}(CP\omega)(k)
 &=\operatorname{Im}\omega(-k^*)
 \leq|\operatorname{Im}(-k^*)|=|\operatorname{Im}k|\,.
 \end{aligned}
 \label{eq:paper-orbit-causality}
\end{equation}
Each transformation is its own inverse, so applying the same argument
to the image proves the converse. This also applies to the smaller
orbit generated by reality alone. Thus, irrespective of the number of
distinct branches, imposing our causality bounds on any one member
of an orbit is necessary and sufficient to impose them on every member
of the orbit.

\subsection{Real dispersion branches}
\label{sec:paper-real-class}

So far nothing was assumed regarding the Taylor coefficients in the expansion \eqref{eq:paper-dispersion-series}. For the rest of the paper, we take both these coefficients to be real: $c_s,\alpha_n\in\mathbb R$. This describes a mode without dissipation. Equivalently, Schwarz reflection gives back the same analytic function,
\begin{equation}
 \omega(k)=[\omega(k^*)]^*.
 \label{eq:paper-schwarz}
\end{equation}
Note that unlike the maps \eqref{eq:paper-branch-maps}, \eqref{eq:paper-schwarz} is a restriction on the functional form of the dispersion relation. In particular, the reality condition of the Green function, \eqref{eq:hermitian-reality}, can exchange branches and is compatible with dissipation. With real coefficients, the orbit \eqref{eq:paper-complex-orbit}, when inversion is a symmetry, reduces to
\begin{equation}
 \{\omega(k),\ \omega(-k),\ -\omega(k),\ -\omega(-k)\}\,.
 \label{eq:paper-real-orbit}
\end{equation}
For a nonzero branch, these are two distinct functions if $\omega(k)$ is purely odd or purely even, and four if both parts are present.

Many familiar examples, are either purely odd or purely even. A sound mode, or the gappless mode in a superfluid is odd. See \cite{Heller:2023hydrohedron,Creminelli:2026dispersion,Esposito:2026diu} for more examples. Gapped modes, such as $\omega=\sqrt{m^2+k^2}$, are even. Examples for mixed dispersion branches include, for example, the two circular polarizations of light in an optically active medium \cite{Landau:1984ecm}, which have
$\omega_\pm=c_sk\pm\alpha_2k^2+\dots$. Another possibility is when there is a preferred direction, such a gapped mode observed in a boosted frame $\omega= m+vk+\dots$.

The main advantage of \eqref{eq:paper-schwarz} is that it simplifies imposing the microcausality condition \eqref{eq:one-sided-causality}.
First, applying that condition at $k^*$ gives
\[
 -\operatorname{Im}\omega(k)=\operatorname{Im}\omega(k^*)
 \leq|\operatorname{Im}k^*|=|\operatorname{Im}k|\,.
\]
Together with \eqref{eq:one-sided-causality} at $k$, this implies the
two-sided bound throughout the disk,
\begin{equation}
|\operatorname{Im}\omega(k)|\leq|\operatorname{Im}k|,
\qquad |k|<R\,.
 \label{eq:two-sided-upper}
\end{equation}
No parity assumption entered this argument. We may therefore constrain
even and odd coefficients together, while treating symmetry restrictions
on the full pole set separately.

We will study the analytic functions satisfying
\eqref{eq:two-sided-upper} for $|k|<R$. This is a necessary condition
for real branches of causal Green functions. Conversely, satisfying
the disk inequality alone is not sufficient for global causality, as a
violation could occur at larger $|k|$. Keeping this distinction explicit
lets us determine sharp coefficient bounds that are optimal given information inside the convergence disk. We discuss separately further restrictions from global continuation outside disk and physical examples.

Schwarz reflection also means that we do not need to check the lower
half of the disk separately: for any point with $\operatorname{Im}k<0$,
its conjugate $k^*$ lies in the upper half, with the same values of
$|\operatorname{Im}\omega|$ and $|\operatorname{Im}k|$. On the upper
half-disk, the two inequalities in \eqref{eq:two-sided-upper} become
\[
 \operatorname{Im}\bigl(k-\omega(k)\bigr)\geq0\,,
 \qquad
 \operatorname{Im}\bigl(k+\omega(k)\bigr)\geq0\,.
\]
Both left-hand sides are \emph{harmonic} functions of
$(\operatorname{Re}k,\operatorname{Im}k)$, since they are imaginary parts
of analytic functions. The minimum principle for harmonic functions
ensures that if they are nonnegative on the boundary of the upper
half-disk, they are nonnegative throughout its interior. For a branch
analytic only in the open disk, we first apply this argument at smaller
radii and then take their limit to $R$.

On the real diameter, both functions vanish by
\eqref{eq:paper-schwarz}, so there is no further condition to impose
there. Hence, it is sufficient to constrain the upper semicircle,
\begin{tcolorbox}[colframe=white,arc=0pt,colback=greyish2]
\vspace{-0.2cm}
\begin{equation}
-\operatorname{Im}k\leq\operatorname{Im}\omega(k)\leq\operatorname{Im}k,
 \qquad k=Re^{i\theta},\quad 0<\theta<\pi\,.
 \label{eq:two-sided-circle}
\end{equation}
\end{tcolorbox}
\noindent Boundary expressions are understood in the limiting sense just described.
As we will discuss in the following section, this problem is equivalent to a one-dimensional moment problem of a positive measure that is bounded from above.

\section{The bounded moment problem}
\label{sec:moment-problem}


We have seen that, assuming analyticity and reality, the microcausality condition on the disc with radius $R$, reduces to a condition over the upper semicircle \eqref{eq:two-sided-circle}. We will encode this condition in a single function bounded between minus one and one, and recover the Taylor coefficients as linear combination of its moments. In this language, bounding a coefficient is a constraint on a moment of a \textit{bounded} function and finding its
largest possible value is a convex optimization problem.

Positive moments have already been used to organize causal coefficient
constraints in the HydroHedron construction
\cite{Heller:2023hydrohedron}. Here the assumption of real Taylor
coefficients gives both an upper and a lower bound on the same boundary
profile. The resulting problem is a weighted version of the bounded
Markov moment problem \cite{KreinNudelman:1977}. We will first establish the connection between bounded moments and dispersion relations in both directions,
then use it to derive bounds.


\subsection{From microcausality to bounded moments}
\label{sec:paper-bounded-moments}

Let $R$ be the radius of a disk centered at $k=0$ on which $\omega(k)$
is analytic and obeys the causal inequality. This radius may be smaller
than the actual convergence radius $R_{\mathrm{conv}}$. We rewrite \eqref{eq:paper-dispersion-series} in terms of the
dimensionless coefficients
\begin{equation}
 a_1=c_s\,,\qquad a_n=\alpha_nR^{n-1}\quad(n\geq2)\,,\qquad
 \omega(k)=R\sum_{n=1}^{\infty}a_n(k/R)^n\,.
 \label{eq:paper-normalized-series}
\end{equation}
Throughout the paper, by Taylor coefficients we mean the coefficients
of positive powers of $k$, namely $a_n$ with $n\geq1$. We drop a possible real gap
$\omega_0$ which does not affect $\operatorname{Im}\omega(k)$
and is therefore not constrained by any of our bounds.
On the upper semicircle, $k=Re^{i\theta}$ with $0<\theta<\pi$,
the real Taylor coefficients give
\begin{equation}
 \operatorname{Im}\omega(Re^{i\theta})
 =R\sum_{n=1}^{\infty}a_n\sin(n\theta)\,.
 \label{eq:paper-boundary-sine-series}
\end{equation}
The causal inequality \eqref{eq:two-sided-circle} bounds $\operatorname{Im}\omega/\operatorname{Im}k$
between $-1$ and $1$ there\footnote{Continuity on the chosen circle is not required. One can first work at radius $R-\epsilon$ and then take $\epsilon\to0$. Values assigned at isolated endpoints or jumps do not affect the moments. We use boundary expressions at $|k|=R$ in this limiting sense.}. Let us define the \textit{profile}
\begin{equation}
 q(\cos\theta):=
 \left[\frac{\operatorname{Im} \omega(k)}{\operatorname{Im}k}\right]_{k=R e^{i\theta}}
 =\frac{\operatorname{Im}\omega(Re^{i\theta})}{R\sin\theta}\,.
 \label{eq:paper-profile}
\end{equation}
The causality condition implies\footnote{\label{qtilde}An equivalent definition would be $2\tilde{q}-1:=\operatorname{Im}\omega/\operatorname{Im}k$ on the upper semicircle. Microcausality condition then implies $0\leq \tilde{q}\leq1$, which has the advantage of being non-negative everywhere. The price is that the expression for $a_1$ acquires an extra constant term. Working with a different normalization and/or a different shift does not change the crucial property we leverage below, namely that $q $ is a bounded function.} 
\begin{tcolorbox}[colframe=white,arc=0pt,colback=greyish2]
\vspace{-0.5cm}
\begin{align}
-1\leq q\leq1\,.
\label{qbound}
\end{align}
\end{tcolorbox}
\noindent To extract the Taylor coefficients, multiply
\eqref{eq:paper-boundary-sine-series} by $\sin(n\theta)$ and use sine
orthogonality:
\begin{equation}
 \begin{aligned}
a_n
 &=\frac2{\pi R}\int_0^\pi
       \operatorname{Im}\omega(Re^{i\theta})\sin(n\theta)\dd{\theta}\\
 &=\frac2\pi\int_0^\pi
       q(\cos\theta)\sin\theta\sin(n\theta)\dd{\theta}\,.
 \end{aligned}
 \label{eq:paper-coefficient-extraction1}
\end{equation}
It is convenient to change the integration variable to $x=\cos\theta$, introduce the normalized semicircle measure, and the Chebyshev polynomials of the second kind,
\begin{equation}
\dd{\mu(x)}:=\frac2\pi\sqrt{1-x^2}\dd{x}\,,\qquad\int_{-1}^1\dd{\mu(x)}=1\,,\qquad
U_{n-1}(\cos\theta):=\frac{\sin(n\theta)}{\sin\theta}\,.
\label{eq:paper-semicircle}
\end{equation}
Then the Taylor coefficients in \eqref{eq:paper-coefficient-extraction1} are written as
\begin{tcolorbox}[colframe=white,arc=0pt,colback=greyish2]
\vspace{-0.1cm}
\begin{equation}
a_n=\int_{-1}^1q(x)U_{n-1}(x)\dd{\mu(x)}\,.
 \label{eq:paper-coefficient-extraction}
\end{equation}
\end{tcolorbox}
\noindent We note that Chebyshev polynomials satisfy orthonormality relation with respect to the normalized measure, i.e. $\int U_{n-1}U_{m-1}\dd{\mu}=\delta_{nm}$. Moreover, the correspondence between the profile and the Taylor coefficients is exact: completeness of the Chebyshev polynomials in $L^2(\mu)$ ensures that all the coefficients determine $q$ uniquely, and vice versa. Since $U_{n-1}(x)$ is a polynomial of degree $n-1$, prescribing the first $N$ Taylor coefficients is equivalent to prescribing the first $N$ moments of $q\dd{\mu}$. For example, 
\begin{equation}
c_s=\int q\dd{\mu}\,,\quad a_2=\int (2x)q\dd{\mu}\,,\quad a_3=\int(4 x^2-1)q\dd{\mu}\,.
\end{equation}

\subsection{Reconstructing the dispersion relation}
\label{sec:paper-profile-reconstruction}
So far, given a causal branch $ \omega(k)$ we have defined a profile $  q(x) $. We can also proceed in the opposite direction: starting from any profile $-1\leq q\leq1$, we can
construct a branch obeying \eqref{eq:two-sided-circle}.

The generating function of the Chebyshev polynomials performs this
construction explicitly. Substituting
\eqref{eq:paper-coefficient-extraction} into
\eqref{eq:paper-normalized-series} and using
\begin{align}
\sum_{n=1}^{\infty}U_{n-1}(x)(k/R)^{n-1}
 =\Big[1-2x(k/R)+(k/R)^2\Big]^{-1}\,,
\end{align} 
gives\footnote{This representation also follows from Cauchy's theorem. Note that the angular average of
$\omega(Re^{i\theta})/[1-(k/R)e^{i\theta}]$ equals $\omega(0)=0$.
Subtracting this average from Cauchy's integral formula and pairing
$\theta$ with $-\theta$ by Schwarz reflection gives
\[
\omega(k)=\frac{2k}{\pi R}\int_0^\pi
\frac{\operatorname{Im}\omega(Re^{i\theta})\sin\theta}
     {1-2(k/R)\cos\theta+(k/R)^2}\,d\theta .
\]
Substituting $\operatorname{Im}\omega(Re^{i\theta})
=R\sin\theta\,q(\cos\theta)$ and setting $x=\cos\theta$ yields
\eqref{eq:paper-profile-reconstruction}.}
\begin{tcolorbox}[colframe=white,arc=0pt,colback=greyish2]
\vspace{-0.1cm}
\begin{equation}
\omega(k)=k\int_{-1}^1
\frac{q(x)\dd{\mu(x)}}{1-2x(k/R)+(k/R)^2}\,,\qquad |k|<R\,.
\label{eq:paper-profile-reconstruction}
\end{equation}
\end{tcolorbox}
\noindent For $x\in[-1,1]$, the denominator can vanish only when $|k|=R$. The integral is therefore analytic inside
the disk, with precisely the real Taylor coefficients prescribed by \eqref{eq:paper-coefficient-extraction}.
We note that \eqref{eq:paper-profile-reconstruction} can also be obtained using the Robertson representation of typically real functions (see \cite{Robertson1935}), which is discussed and used by \cite{Creminelli:2026dispersion}. 

As mentioned before, by the maximum principle for harmonic functions, imposing the microcausality condition on the boundary, i.e. $-1\leq q\leq1$, ensures that it is satisfied within the whole disc. This can be verified directly from \eqref{eq:paper-profile-reconstruction}.  Taking the imaginary part of the kernel
in \eqref{eq:paper-profile-reconstruction} gives,
\begin{equation}
\operatorname{Im}\omega(k)
=\operatorname{Im}k\int_{-1}^1
\frac{1-|k|^2/R^2}{|1-2x(k/R)+(k/R)^2|^2}q(x)\dd{\mu(x)}\,.
\label{eq:paper-reconstruction-average}
\end{equation} 
The kernel is non-negative for \(|k|<R\) and integrates to one,
as follows by taking \(q(x)=1\), the profile of \(\omega(k)=k\):
\begin{align}\label{tobeadded}
 \int_{-1}^1
 \frac{1-|k|^2/R^2}{|1-2x(k/R)+(k/R)^2|^2}\,\dd{\mu(x)}=1.
\end{align}
Thus the integral in \eqref{eq:paper-reconstruction-average} is a
weighted average of \(q\). Since \(-1\leq q\leq1\), it follows that \(|\operatorname{Im}\omega(k)|\leq|\operatorname{Im}k|\) as required.


We anticipate that the radius of convergence of $\omega(k)$, as given in \eqref{eq:paper-profile-reconstruction}, may be larger than $R$, as we will discuss in Section~\ref{sec:convergence-radius}.


\subsection{Markov vs. Hausdorff moment problem}\label{markov}

The Taylor coefficients are linear combinations of the moments of $q(x)\dd{\mu(x)}$, subject to \eqref{qbound}. To compare with the moment problems that are more familiar, it is convenient to shift and rescale, $\tilde q:=(q+1)/2$, so that $0\le\tilde q\le1$ (see footnote~\ref{qtilde}). The moments of a density bounded from above \emph{and} below
define the Markov moment problem, or $L$-problem
\cite{AkhiezerKrein1962,KreinNudelman:1977,diaconis2004markov1}. Before
studying \eqref{eq:paper-coefficient-extraction}, it is instructive to see in the simplest setting the implications of a point-wise upper bound on the measure. Consider,
\begin{equation}
m_j=\int_a^b x^j \tilde{q}(x)\dd{x} ,\quad 0\le \tilde{q}(x)\le 1\,, \quad j=0,1,\cdots\,,
\label{eq:markov-moments}
\end{equation}
where for simplicity we have replaced ${\rm d}\mu$ by ${\rm d}x$ on a generic interval $[a,b]$. If only $\tilde q\ge0$ is imposed, this is the Hausdorff moment problem \cite{Akhiezer1965,schmudgen2017moment}, which underlies positivity bounds on scattering amplitudes \cite{Bellazzini:2020cot}. See also \cite{Apostolidis:2026mvb}. The bounded problem differs from it in at least three ways. 

First of all, the allowed region of moments $\{m_j\}^{\infty}_{j=0}$ is compact rather than a cone. If only $\tilde q\ge0$ is required, $\lambda\tilde q$ is allowed for every $\lambda>0$. The moments can therefore be rescaled, and only their ratios are constrained, i.e. the allowed region is a cone. The upper bound removes this freedom and each moment is bounded. For instance from \eqref{eq:markov-moments} we see that $0\le m_0\le b-a$.  The allowed region is therefore compact.

Secondly, there are actually two positive measures rather than one. This implies that, sending $\tilde{q}\to1-\tilde{q}$, both $\tilde{q}$ and $1-\tilde{q}$ are non-negative. Therefore, both $\{m_j\}$ and the complementary
moments $\{(b^{j+1}-a^{j+1})/(j+1)-m_j\}$ obey the Hankel-matrix conditions of the Hausdorff problem \cite{Bellazzini:2020cot,schmudgen2017moment}. The allowed region is then symmetric under their exchange. Truncated at finite order, these two sets of Hankel-matrix conditions are necessary but not sufficient; they become sufficient only as the order goes to infinity. 

Finally, the boundary of the allowed region is reached by step functions rather than delta functions. In the Hausdorff problem the boundary of the allowed region is reached by concentrating the measure $\tilde{q}(x)$ into a sum of delta functions, known also as atomic measures. These are now forbidden, since they violate $\tilde q\le1$. The most concentrated allowed density is $\tilde q=1$ on an interval, and the boundary is reached by densities that alternate between $0$ and $1$ a finite number of times. In both problems a boundary point determines its density uniquely.

\begin{figure}
\centering
\includegraphics[width=0.45\linewidth]{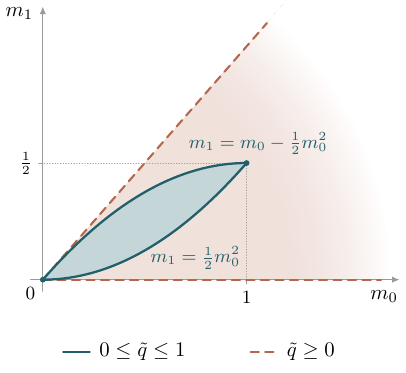}
\caption{The allowed region for $\{m_0,m_1\}$ moments of $\tilde{q}$ over the interval $[0,1]$. The comparison is between the cases with or without an upper bound. In the latter, the region is a convex cone (the red shade). The former is a convex bounded region, which is symmetric under $\{m_0,m_1\}\to\{1-m_0,\frac{1}{2}-m_1\}$. The boundary of the cone is reached by $\tilde{q}=m_0\delta(x)$ and $\tilde{q}=m_0\delta(x-1)$. The boundary of the bounded region is reached step functions, i.e. $\tilde{q}=\Theta(m_0-x)$ on the lower curve and $\tilde{q}=\Theta(x-1+m_0)$ on the upper one. See section \ref{sec:paper-threshold-profiles}.}
\label{fig:markov}
\end{figure}

The comparison is shown in Fig.~\ref{fig:markov} for a simple case of the first two moments $\{m_0,m_1\}$ over the interval $[0,1]$. 

Most of the above lessons are transferred to the case where we work directly with $-1\leq q\leq1$ and with $\dd{\mu}$ instead of $\dd{x}$. This will be the topic of the following sections. An important property which will be used in the following is that the allowed region of moments, in all cases, is a convex set. If $q_1$ and $q_2$ are allowed, i.e.~satisfy $  -1\leq q_{1,2}\leq 1 $, so is $\lambda q_1+(1-\lambda)q_2$ for any $0\leq \lambda\leq1$. All the
coefficients in \eqref{eq:paper-coefficient-extraction} transform by the
same convex combination.  In terms of the dispersion relations, the
reason is simply
\begin{equation}
|\operatorname{Im}[\lambda\omega_1(k)+(1-\lambda)\omega_2(k)]|
 \leq \lambda|\operatorname{Im}\omega_1(k)|
       +(1-\lambda)|\operatorname{Im}\omega_2(k)|
 \leq|\operatorname{Im}k|\,.
 \label{eq:paper-common-radius-convexity}
\end{equation}
Therefore, every finite projection, such as $(c_s,a_2,\ldots,a_N)$, is also a convex set. Our strategy to find the allowed region is simply to obtain its boundary since the interior will be automatic, as a consequence of convexity. In the following section, we will see how to obtain the unique profile $q$ corresponding to the boundaries.  


\subsection{Extremal profiles}
\label{sec:paper-threshold-profiles}


We start with two examples that maximize a Taylor
coefficient and compare them with a generic allowed profile. This also
shows how maximizing one coefficient fixes the profile and hence all
the remaining coefficients.

Let us maximize $a_2$, allowing all other coefficients,
including $c_s$, to vary. Since $a_2=\int (2x)q(x)\dd{\mu(x)}$, we should set
$q=-1$ where $x<0$ to avoid a negative contribution, and $q=1$ where
$x>0$ to make the positive contribution as large as possible. Thus
\begin{equation}
 q(x)=\operatorname{sgn}(x),\quad\implies\quad
 a_2=2\int_0^1(2x)\dd{\mu(x)}=\frac8{3\pi}\,.
 \label{eq:paper-quadratic-profile-example}
\end{equation}
This example also explains why saturating one bound fixes the other
coefficients. Adding weight at $x<0$ or removing it at $x>0$ would decrease
$a_2$. The value of $q$ at $x=0$ does not affect the integrals, so there is no
remaining freedom in any moment. Substitution into
\eqref{eq:paper-coefficient-extraction} gives, for example,
$c_s=a_3=0$, $a_4=-16/(15\pi)$, and $a_6=24/(35\pi)$.
All subsequent coefficients are fixed in the same way.

The same reasoning applies to $a_3=\int(4x^2-1)q(x)\dd{\mu(x)}$.
The polynomial is positive for $|x|>1/2$ and negative for $|x|<1/2$,
so the maximizing profile is
\begin{equation}
 q(x)=\operatorname{sgn}(U_2(x))=\operatorname{sgn}(4x^2-1),\quad\implies\quad
 a_3=\frac{3\sqrt3}{2\pi}\,.
 \label{eq:paper-cubic-profile-example}
\end{equation}
Once again, all the other Taylor coefficients are fixed as well. Equivalently, \eqref{eq:paper-profile-reconstruction} fixes the dispersion relation. We will give the resulting dispersion
relations in Section~\ref{sec:paper-individual-bounds}.

\begin{figure}[tb]
 \centering
 \includegraphics[width=\textwidth]{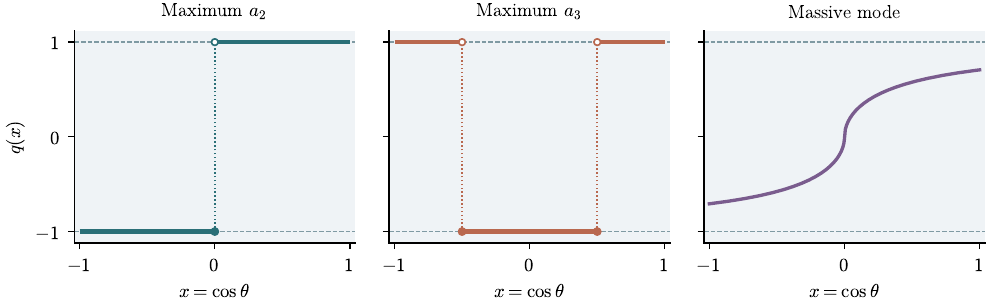}
 \caption{Two extremal profiles and a non-extremal massive mode. Left:
 $q=\operatorname{sgn}(x)$ attains $a_2=8/(3\pi)\simeq0.849$.
 Middle: $q=\operatorname{sgn}(4x^2-1)$ attains $a_3=3\sqrt3/(2\pi)\simeq0.827$.
 Right: the relativistic branch $\omega=\sqrt{m^2+k^2}$ at $R=m$ gives
 a continuous allowed profile. In the two maximizations,
 all other coefficients were initially free. Values at the jumps do not
 affect any coefficient.}
\label{fig:paper-profile-examples}
\end{figure}

Indeed a generic allowed profile need not take only the values $-1$ and $1$.
For comparison, consider the massive relativistic branch
$\omega(k)=\sqrt{m^2+k^2}$ with $m>0$. At its true convergence radius
$R=m$, \eqref{eq:paper-profile} gives
\begin{equation}
 q(x)=\operatorname{sgn}(x)
                    \sqrt{\frac{|x|}{1+|x|}}\,.
 \label{eq:paper-massive-profile-example}
\end{equation}
This is continuous, but not differentiable at $x=0$, profile which stays strictly between minus one and one. It has
$c_s=0$, like the quadratic extremizer, but $a_2=1/2<8/(3\pi)$.
Figure~\ref{fig:paper-profile-examples} compares the three profiles.

These examples generalize directly. To maximize any $a_n$ without fixing
any other Taylor coefficient: set $q=1$ where $U_{n-1}>0$ and $q=-1$
where $U_{n-1}<0$. Similarly, for any nonzero linear combination of Taylor coefficients, we construct the corresponding linear combination of Chebyshev polynomial, and the extremal profile is fixed by its sign. We use this
construction below to derive bounds on individual coefficients and
their linear combinations.


\section{Linear bounds}
\label{sec:linear-bounds}

Our reformulation of causality constraints into moments of a bounded function turns extremization of a Taylor coefficient into the choice of where the profile $q$  should be largest, $q=1$, or smallest, $q=-1$. We use this to find the largest and smallest allowed values of the Taylor coefficients, first for individual coefficients and then for arbitrary linear combinations. Throughout, we use the common radius $R$ and the real coefficients $a_1=c_s$ and $a_n=\alpha_nR^{n-1}$. In this section, when a linear combination of coefficients is extremized, all the others are varied freely.


\subsection{Bounds on individual Taylor coefficients}
\label{sec:paper-slope-bound}
\label{sec:paper-individual-bounds}
\label{sec:paper-saturation-slopes}

Let us start finding bounds for the first Taylor coefficient, namely the speed of sound $  c_{s} $. From the normalization condition \eqref{eq:paper-semicircle}, and \eqref{eq:paper-coefficient-extraction} we immediately obtain
\begin{equation}
c_s=\int_{-1}^1q(x)\dd{\mu(x)}\,,\quad\implies\quad
-1\leq c_s\leq1\,.
\label{eq:paper-slope-bound}
\end{equation}
where the bound follows from using $-1\leq q\leq 1$. Every $c_s$ strictly inside this interval can be realized by infinitely
many profiles $  q $. At the maximum (minimum), however, a constant profile $q=1$ ($q=-1$) is the only possibility. For these extrema all higher moments vanish by orthonormality of the Chebyshev polynomials
\begin{tcolorbox}[colframe=white,arc=0pt,colback=greyish2]
\vspace{-0.1cm}
\begin{equation}
c_s=\pm1\quad\Longrightarrow\quad
 a_n=0\ \ (n\geq2),\qquad \omega(k)=\pm k\,.
 \label{eq:paper-endpoint-rigidity}
\end{equation}
\end{tcolorbox}
\noindent This recovers the known bound on $c_s$ and the rigidity of the luminal
branch, established in Ref.~\cite{Heller:2022ejw,Heller:2023hydrohedron} and discussed in
Ref.~\cite{Creminelli:2026dispersion}.

\begin{figure}[tb]
 \centering
 \includegraphics[width=.7\textwidth]{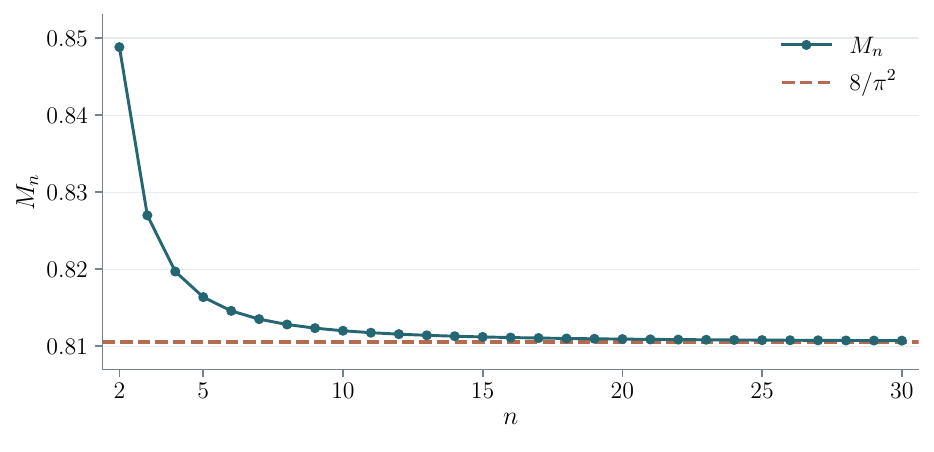}
 \caption{The sharp maximum $M_n$ of each Taylor coefficient $a_n$, for
 integer orders $2\leq n\leq30$. The dashed line marks the asymptote
 $\lim_{n\to\infty}M_n=8/\pi^2\simeq0.81057$.
 The minimum at each order is $-M_n$.}
 \label{fig:paper-individual-bounds}
\end{figure}

We now derive the profiles that extremize each Taylor coefficient $a_n$ individually, for $n\geq2$. As in Section~\ref{sec:paper-threshold-profiles},
the extremizing profiles are
\begin{equation}
 q=\operatorname{sgn}(U_{n-1})\quad\text{at the maximum}\,,\qquad
 q=-\operatorname{sgn}(U_{n-1})\quad\text{at the minimum}\,.
 \label{eq:paper-individual-extremizers}
\end{equation}
Flipping the profile reverses every coefficient, so the lower bound
has the same magnitude. Substituting the maximizing profile into
\eqref{eq:paper-coefficient-extraction} and using
\begin{align}
\operatorname{sgn}(U_{n-1})U_{n-1}=|U_{n-1}|\,,
\end{align} 
gives the sharp bounds\footnote{To compute the integral, split it at the
zeros of $\sin(n\theta)$ and use the finite sine sum in
ref.~\cite[eq.~(21)]{Weisstein:Sine}.}
\begin{tcolorbox}[colframe=white,arc=0pt,colback=greyish2]
\vspace{-0.0cm}
\begin{equation}
|a_n|\leq M_n:=\frac2\pi\int_0^\pi|\sin(n\theta)|\sin\theta\,d\theta=\frac{4n}{\pi(n^2-1)}\cot\frac\pi{2n},
 \qquad n\geq2\,.
 \label{eq:paper-individual-bound}
\end{equation}
\end{tcolorbox}
\noindent For example, their first two values, consistent with \eqref{eq:paper-quadratic-profile-example} and \eqref{eq:paper-cubic-profile-example}, and their
large-order limit are
\begin{equation}
 M_2=\frac8{3\pi},\qquad M_3=\frac{3\sqrt3}{2\pi},\qquad
 \lim_{n\to\infty}M_n=\frac8{\pi^2}\,.
 \label{eq:paper-individual-examples}
\end{equation}
In particular, unlike $  c_{s} $, every higher-order coefficient $  a_{n} $ for $  n\geq 2 $ has a maximal magnitude
strictly below one. Figure~\ref{fig:paper-individual-bounds} shows the individual maxima and
their large-order asymptote. Moreover, at extrema of $a_n$ for $n\geq 2$, infinitely many other Taylor coefficients are nonzero. 

\subsubsection*{Extremal dispersion relations} 

Saturation of a bound  on one $  a_{n} $ fixes the full dispersion relation or equivalently all the Taylor coefficients. We can therefore reconstruct the extremal dispersion relation for each of them. Denote by $\omega_n^{\rm max}(k)$ the branch maximizing $a_n$.
On the angular intervals $j\pi/n<\theta<(j+1)\pi/n$, its profile obeys
$q(\cos\theta)=\operatorname{sgn}(\sin n\theta)=(-1)^j$.
Substituting this into \eqref{eq:paper-profile-reconstruction} gives
\begin{equation}
 \omega_n^{\rm max}(k)=\frac{2k}{\pi}
 \sum_{j=0}^{n-1}(-1)^j
 \int_{j\pi/n}^{(j+1)\pi/n}
 \frac{\sin^2\theta\dd{\theta}}
 {1-2(k/R)\cos\theta+(k/R)^2}\,.
 \label{eq:paper-extremal-dispersion-integral}
\end{equation}
Each integral is elementary and evaluating the result at the endpoints yields the
closed expression\footnote{For odd $n\geq3$, the maximizing dispersion
relation takes the somewhat simpler form
\[
 \begin{aligned}
 \omega_n^{\rm max}(k)=\frac{k}{n}+\frac{2R}{\pi}\left(\frac{k}{R}-\frac{R}{k}\right)
 \sum_{j=1}^{(n-1)/2}(-1)^{j-1}
 \arctan\frac{(k/R)^2\sin(2j\pi/n)}
 {1-(k/R)^2\cos(2j\pi/n)}.
 \end{aligned}
\]
This follows by pairing the terms $j$ and $n-j$. The arctangents are
continued from $k=0$.}
\begin{tcolorbox}[colframe=white,arc=0pt,colback=greyish2]
\vspace{-0.0cm}
\begin{align}\label{eq:paper-extremal-dispersion}
 \omega_n^{\rm max}(k)=&\frac{1-(-1)^n}{2n}k\\
 &+\frac{2R}{\pi}\sum_{j=1}^{n-1}(-1)^{j-1}
 \left[\sin\frac{j\pi}{n}
 +\left(\frac{k}{R}-\frac{R}{k}\right) \nonumber
 \arctan\frac{(k/R)\sin(j\pi/n)}{1-(k/R)\cos(j\pi/n)}\right]\,.
\end{align}
\end{tcolorbox}
\noindent The constants $\sin(j\pi/n)$ in the second line, ensure that the sum does not contribute to a gap, i.e. $\omega_n^{\rm max}(0)=0$, or in case we include a real gap $\omega_n^{\rm max}(0)=\omega_0$. The apparent singularity at $  k=0 $ is absent once expanding the full expression. This expression is analytic
and satisfies the causal inequality for $|k|<R$. The minimizing branch is obtained by $q(x)\to-q(x)$, 
\begin{equation}
 \omega_n^{\rm min}(k)=-\omega_n^{\rm max}(k)\,.
 \label{eq:paper-minimal-dispersion}
\end{equation}
For $n=2$, the sum in \eqref{eq:paper-extremal-dispersion} contains only a single term and reduces to
\begin{equation}
 \omega_2^{\rm max}(k)=\frac{2R}{\pi}
 \left[1+\left(\frac{k}{R}-\frac{R}{k}\right)
 \arctan\frac{k}{R}\right]\,.
 \label{eq:paper-quadratic-extremal-dispersion}
\end{equation}
Expanding the arctangent in \eqref{eq:paper-quadratic-extremal-dispersion} fixes every coefficient explicitly: 
\begin{equation}
 a_{2j}=\frac{8j(-1)^{j-1}}{\pi(4j^2-1)}\quad(j\geq1)\,,
 \qquad a_{2j+1}=0\quad(j\geq0)\,.
 \label{eq:paper-quadratic-extremal-coefficients}
\end{equation}
This reproduces $a_2=8/(3\pi)$, $a_4=-16/(15\pi)$, and
$a_6=24/(35\pi)$ from Section~\ref{sec:paper-threshold-profiles}.
The branch points at $k=\pm iR$ give its true convergence radius $R$.
For general $n\geq2$, the arctangents in
\eqref{eq:paper-extremal-dispersion} have branch points at
$k=R e^{\pm i j\pi/n}$, $j=1,\ldots,n-1$, so the radius of convergence is again $ R $. We return to this in Section~\ref{sec:convergence-radius}. These branches show that the bounds \eqref{eq:paper-individual-bound} are sharp: they cannot
be improved using only analyticity and causality on the disk. Whether an
extremal branch can be a pole of a causal Green function beyond the disk is
a separate question, to which we return in Section~\ref{extdisp}.


\subsection{Arbitrary linear combinations and completeness}
\label{sec:paper-linear-combinations}

Separate bounds on each coefficient do not tell us which values can occur
together. To describe this, let $\mathcal{A}_N\subset\mathbb{R}^N$ be the
set of values of $(a_1,\dots,a_N)$ produced by some profile with
$|q|\le1$, all higher coefficients being free. We saw in
Section~\ref{markov} that $\mathcal{A}_N$ is convex, and it is
compact. A compact convex set is the intersection of
the half-spaces that contain it. It is therefore completely specified by
the largest value of $\sum_n t_na_n$ in every direction
$t=(t_1,\dots,t_N)$. For any nonzero linear combination of Taylor coefficients, we can construct the polynomial
\begin{equation}
 P_t(x)=\sum_{n=1}^N t_nU_{n-1}(x)\,,\qquad
\sum_{n=1}^N t_na_n=\int_{-1}^1q(x)P_t(x)\dd{\mu(x)}\,.
 \label{eq:paper-moment-objective}
\end{equation}
Since $|q(x)|\leq1$, we obtain the sharp family of bounds
\begin{equation}
\left|\sum_{n=1}^N t_na_n\right|
 \leq\int_{-1}^1|P_t(x)|\dd{\mu(x)}\,.
 \label{eq:paper-linear-support}
\end{equation}
The upper extremum is attained by $q=\operatorname{sgn}(P_t)$ and the lower
one by $q=-\operatorname{sgn}(P_t)$. The individual bounds derived previously correspond to a single nonzero
$t_n$. Note that, a point $(a_1,\dots,a_N)$ belongs to $\mathcal{A}_N$ if and only if it satisfies \eqref{eq:paper-linear-support} for every $t$, because a point outside a compact convex set is separated from it by a hyperplane. The linear bounds are thus necessary and sufficient, and all other bounds in this paper are consequences of them. Moreover, the profile saturating a given bound is unique. Therefore, saturation fixes all Taylor coefficients, and hence the full dispersion relation.

For nonzero $P_t$, let $x_1<\dots<x_J$ be the points in $(-1,1)$ where it
changes sign, and set $x_0=-1$ and $x_{J+1}=1$. Since $P_t$ has a definite
sign between consecutive points, the integrated form of the bound is
\begin{tcolorbox}[colframe=white,arc=0pt,colback=greyish2]
\vspace{-0.0cm}
\begin{equation}
\left|\sum_{n=1}^N t_na_n\right|
\leq\sum_{j=0}^{J}\left|K_t(x_{j+1})-K_t(x_j)\right|\,.
 \label{eq:paper-integrated-linear-support}
\end{equation}
\end{tcolorbox}
\noindent Here $K_t(x):=\sum_{n=1}^{N}t_nK_n(x)$ is built from the cumulative measure
and the cumulative Chebyshev moments, which we will use also in Section~\ref{sec:nonlinear-bounds}:\footnote{We simplified the expression for $n\geq2$ using $(1-x^2)U_{n-1}''-3xU_{n-1}'+(n^2-1)U_{n-1}=0$.}
\begin{equation}
\begin{aligned}
  K_1(x)&:=\int_{-1}^{x}\mathrm{d}\mu
       =\frac12+\frac1\pi\Big[\arcsin x+x\sqrt{1-x^2}\,\Big]\,,\\
  K_n(x)&:=\int_{-1}^{x}U_{n-1}\,\mathrm{d}\mu
       =-\frac{2\,(1-x^2)^{3/2}}{\pi(n^2-1)}\,U_{n-1}'(x)\,,
       \qquad n\ge2\,.
\end{aligned}
\label{eq:paper-linear-support-antiderivative}
\end{equation}
so that $K_n(\pm1)=0$ for $n\ge2$, while $K_1(-1)=0$ and $K_1(1)=1$. Special cases, which will be used later, are
\begin{equation}
K_2(x)=-\frac{4}{3\pi}(1-x^2)^{3/2}\,,\qquad
K_3(x)=-\frac{2}{\pi}\,x\,(1-x^2)^{3/2}\,.
\end{equation}
We emphasize that \eqref{eq:paper-integrated-linear-support} identifies the profiles on the boundary of $\mathcal{A}_N$. Every boundary point of a convex set saturates the bound in some direction $t\neq0$, so its profile is $q=\operatorname{sgn}P_t$. This profile takes only the values $\pm1$, and it changes sign at most $N-1$ times because $P_t$ has degree at most $N-1$. Conversely, a profile taking values $\pm1$, with $J\le N-1$ sign changes at $x_1,\dots,x_J$, is given by $\pm\operatorname{sgn}\prod_j(x-x_j)$. It is therefore the sign of a polynomial of the form $P_t$ with degree at most $N-1$, so it lies on the boundary.

\begin{figure}[tb]
 \centering
 \includegraphics[width=\textwidth]{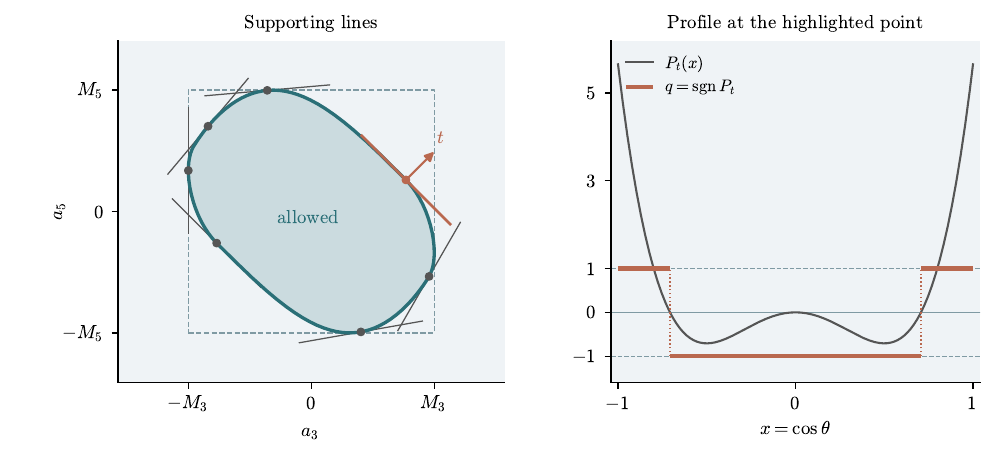}
 \caption{Linear bounds as supporting lines. Left: the exact $(a_3,a_5)$ region (shaded), with all
other Taylor coefficients free. Each direction $t=(t_3,t_5)$ gives one line
$t_3a_3+t_5a_5=\int|P_t|\,d\mu$ with $P_t=t_3U_2+t_5U_4$, which touches the region at a single
point; the region is the intersection of the corresponding half-planes over all directions.
The dashed rectangle shows the individual bounds $|a_3|\le M_3$ and $|a_5|\le M_5$. The
highlighted line has $t_3=t_5$ and is the bound $a_3+a_5\le 8/(3\pi)$ of \eqref{eq:paper-cubic-quintic-sum}, touching at
$(a_3,a_5)=(2/\pi,\,2/(3\pi))$. Right: the polynomial $P_t$ for the highlighted direction,
normalized to $|t|=1$, and the profile $q=\operatorname{sgn}(P_t)=\operatorname{sgn}(|x|-1/\sqrt{2})$
that attains the bound.}
 \label{fig:supporting}
\end{figure}

We finish this section by an example. Take $t_3=t_5=1$ and the other weights zero gives
$P_t=U_2+U_4=8x^2(2x^2-1)$. Hence
\begin{equation}
|a_3+a_5|\leq\frac8{3\pi}\,.
\label{eq:paper-cubic-quintic-sum}
\end{equation}
We see that this simple joint bound already improves substantially on adding individual
ones, i.e. $M_3+M_5\simeq1.64$. The maximizing profile is $q=\operatorname{sgn}(|x|-1/\sqrt2)$. Its moments include
\begin{equation}
 c_s=-\frac2\pi\,,\qquad a_3=\frac2\pi\,,\qquad
 a_5=\frac2{3\pi}\,,
 \label{eq:paper-cubic-quintic-extremizer}
\end{equation}
which shows directly the bound is saturated. See Figure~\ref{fig:supporting}.

\section{Nonlinear bounds}
\label{sec:nonlinear-bounds}

The linear bounds of Section~\ref{sec:linear-bounds} characterize every finite coefficient region, but using them directly requires testing infinitely many linear combinations. In some cases, a nonlinear inequality can package these tests into
a condition on a few coefficients we know. This may be more useful to a low-energy observer measuring just a few low-order Taylor coefficients. In this section we begin deriving a useful sphere bound involving all Taylor coefficients. Then we give exact descriptions of a few two- and three-coefficient regions. Coefficients that are not displayed are varied freely.


\subsection{The sphere bound}
\label{sec:paper-parseval}

The key idea to derive the sphere bound is to use Parseval's theorem. The coefficient formula \eqref{eq:paper-coefficient-extraction} is an
orthogonal expansion of $q(x)$ in the Chebyshev basis. Since $q$ is
bounded, this expansion converges:
\begin{equation}
 q(x)=\sum_{n=1}^{\infty}a_nU_{n-1}(x)\,.
 \label{eq:paper-parseval-expansion}
\end{equation}
If we use the fact that orthogonality removes cross terms when we integrate the square of a
finite partial sum we find
\begin{equation}
 \sum_{n=1}^{\infty}a_n^2
 =\int_{-1}^1q(x)^2\dd{\mu(x)}
 =1-\int_{-1}^1\big[1-q(x)^2\big]\dd{\mu(x)}\leq1\,.
 \label{eq:paper-parseval-identity}
\end{equation}
The infinite sum captures the full squared norm so this is just Parseval's theorem. In the final step we use the causality condition $|q|\leq1$ to obtain the sphere bound
\begin{tcolorbox}[colframe=white,arc=0pt,colback=greyish2]
\vspace{-0.0cm}
\begin{equation}
\sum_{n=1}^{\infty}a_n^2\leq1\,.
\label{eq:paper-parseval-bound}
\end{equation}
\end{tcolorbox}
\noindent This bound is sharp: it is saturated precisely when $q$ takes only the values $\pm1$ almost everywhere, as evident from the second equality in \eqref{eq:paper-parseval-identity}. Thus each of the extremizing dispersion relations discussed in the previous section saturates it. The sphere bound is only necessary but not sufficient for causality: for example, $a_2=1$ obeys
the sphere bound if all other coefficients vanish but violates
\eqref{eq:paper-individual-bound}. The reason is that \eqref{eq:paper-parseval-bound} only constrains
the average of $q^2$ over the semicircle, whereas causality requires $|q|\le1$ pointwise. To bring this point home, an analogy may be useful. Consider a 2d ball in 3 dimensions intersected with a 2d paraboloid. Allowed points must be both inside the ball and inside the paraboloid; however, being inside one of them is not sufficient. 

There is also a projected sphere bound that remains sharp, i.e.~it can be saturated. If $a_1=\cdots=a_m=0$ for any finite $m\geq1$, the sphere bound becomes
\begin{equation}
 \sum_{n=m+1}^{\infty}a_n^2\leq1-\sum_{n=1}^m a_n^2=1\,.
 \label{eq:paper-vanishing-tail}
\end{equation}
This can be saturated and indeed an explicit saturating profile is
\begin{equation}
q(\cos\theta)=\operatorname{sgn}\Big(\cos\bigl((m+2)\theta\bigr)\Big)\,.
\label{eq:paper-vanishing-profile}
\end{equation}
The square wave $q(\cos\theta)$ has cosine harmonics only at odd multiples of
$m+2$. In the moment integral \eqref{eq:paper-coefficient-extraction1}, the identity
$2\sin\theta\sin(n\theta)=\cos((n-1)\theta)-\cos((n+1)\theta)$
then makes the first $m$ coefficients vanish. Since the profile takes values that are either one or minus one, it saturates the sphere bound. Note that the sphere bound cannot be saturated with finitely many nonzero coefficients, except by $\omega=\pm k$: saturation requires $|q|=1$ almost
everywhere, and a polynomial $q$ with this property is constant. In particular, no truncation $\sum_{n\le N}a_n^2\le1$ with $N\ge2$ is sharp away from $c_s=\pm1$.


\subsection{Fixing a subset of coefficients}
\label{sec:fixing}

Section~\ref{sec:paper-linear-combinations} gives the bound on a combination
$\sum_n t_n a_n$ when all coefficients are free. In practice however one often knows only some coefficients, for instance $c_s$, and one wants the sharp range of another one \emph{given} the known information. The constrained problem can be solved by introducing a Lagrange multiplier for each fixed coefficient and using techniques similar to the previous section.

Let $S$ be a finite set of Taylor coefficients, and assume the values
$a_m=\bar a_m$ for $m\in S$ are given and fixed. We want to find the maximum of some finite combination of the remaining coefficients, $\sum_{n\notin S} t_n a_n$, over all admissible profiles with the prescribed moments:
\begin{equation}
B(\bar a):=\max_{q}\ \int_{-1}^{1}q(x)P_t(x)\dd{\mu(x)}\,,
  \quad\ \text{subject to}\quad\
  |q|\le 1\,,
  \quad
  \bar a_m=\int_{-1}^{1} q\,U_{m-1}\dd{\mu}\,,
  \label{eq:constrained}
\end{equation}
for $m\in S$, and $P_t(x)=\sum_{n\notin S} t_n\,U_{n-1}(x)$. As discussed in Section \ref{sec:paper-linear-combinations}, without the prescribed moments, the maximization is immediate since the condition $|q(x)|\le 1$ can be imposed at each point, and different points do not talk to each other. Moment constraints spoil this property, and we need to introduce Lagrange multipliers. 

For any profile obeying the constraints, the combination
$\sum_{m\in S}\lambda_m\big(\bar a_m-\int q\,U_{m-1}\dd\mu\big)$ vanishes, with arbitrary $\lambda_m$. Therefore, adding it to the quantity we want to maximize is harmless:
\begin{equation}
  \int_{-1}^{1} q\,P_t\dd\mu
  =\sum_{m\in S}\lambda_m\bar a_m+\int_{-1}^{1} q\,Q_\lambda\dd\mu\leq\sum_{m\in S}\lambda_m\bar a_m+\int_{-1}^{1}\big|Q_\lambda\big|\dd\mu\,,
  \label{eq:addzero}
\end{equation}
where
\begin{equation}
Q_\lambda(x):=P_t(x)-\sum_{m\in S}\lambda_m\,U_{m-1}(x)\,.
\end{equation}
The last inequality in \eqref{eq:addzero} is obtained by noting that the integral is at most $\int|Q_\lambda|\,d\mu$ for \emph{every} profile with $|q|\le1$, whether or not it satisfies the moment constraints. The best bound is obtained by minimizing over $\lambda$,\footnote{The steps from \eqref{eq:constrained} to \eqref{eq:strongduality} are an instance of Lagrangian duality for a linear program \cite{Luenberger1969,BoydVandenberghe2004}. In the
\emph{primal} problem \eqref{eq:constrained} one maximizes over profiles, so any admissible
profile gives a lower bound on $B(\bar a)$. In the \emph{dual} problem \eqref{eq:strongduality} one minimizes over multipliers, so any $\lambda$ gives a rigorous upper bound. Equation~(\ref{eq:addzero}) is weak duality. The fact that the two optima coincide is known as strong duality. The same primal--dual structure
underlies the dual formulations of the S-matrix and EFT bootstrap \cite{Caron-Huot:2020cmc}.} 
\begin{tcolorbox}[colframe=white,arc=0pt,colback=greyish2]
\vspace{-0.0cm}
\begin{equation}
B(\bar a)=\min_{\lambda}\,\bigg\{\sum_{m\in S}\lambda_m\bar a_m
  +\int_{-1}^{1}\Big|P_t-\sum_{m\in S}\lambda_m U_{m-1}\Big|\dd\mu\bigg\}\,.
  \label{eq:strongduality}
\end{equation}
\end{tcolorbox}
\noindent We note that the inequality in \eqref{eq:addzero} is saturated only if $q=\operatorname{sgn}Q_\lambda$ almost everywhere. One can show that at the stationary point, i.e. in \eqref{eq:strongduality}, the profile is automatically consistent with the moment constraints, e.g. by varying the right hand side of \eqref{eq:addzero} with respect to $\lambda$. We conclude that the extremal profile of \eqref{eq:strongduality} is given by
\begin{equation}
q_{\max}=\operatorname{sgn}Q_{\lambda_*}=\operatorname{sgn}\Big(P_t-\sum_{m\in S}\lambda_{*m} U_{m-1}\Big)\,,
\label{extq}
\end{equation}
where $\lambda_*$ is the stationary point in \eqref{eq:strongduality}. Note that $Q_\lambda$ has the same form as \eqref{eq:paper-moment-objective}, with weights $t_n$ for $n\notin S$ and $-\lambda_m$ for $m\in S$. Fixing some coefficients thus selects, among the linear bounds \eqref{eq:paper-linear-support}, the one that is saturated at the prescribed values. In particular all the conclusions of Section~\ref{sec:paper-linear-combinations} carry over: the extremal profile takes only the values $\pm1$, it is unique, and it fixes all the other Taylor coefficients.

Finally, under $q\to-q$ all coefficients change sign. The smallest value of the linear combination, given $\bar a$, is therefore $-B(-\bar a)$. Since the constraints are linear in $q$, the constrained problem is still convex with respect to linear combination of different profiles. As a result, all values between $-B(-\bar a)$ and $B(\bar a)$ are admissible, and the allowed region is a convex set. 

Let us see how this works when only
$c_s=a_1$ is prescribed. Since $U_0=1$, the constraint is $\int q\dd\mu=c_s$, there is a single multiplier, and the shifted polynomial is $Q_\lambda=P_t-\lambda$. Equation
\eqref{eq:strongduality} becomes
\begin{equation}
  \max_{q:\ \int q\,d\mu=c_s}\ \sum_{n\ge2}t_na_n
  =\min_{\lambda}\bigg\{\lambda c_s+\int_{-1}^{1}|P_t-\lambda|\,d\mu\bigg\}\,,
  \qquad
  q_{\max}=\operatorname{sgn}(P_t-\lambda_*)\,.
  \label{eq:fixedcs}
\end{equation}
The Lagrange multiplier has a simple interpretation as threshold here: we set $q=+1$ only where $P_t$ exceeds $\lambda$, and tune
$\lambda$ until the profile has the required average given by $c_s$:
\begin{equation}
c_s(\lambda)=\int_{-1}^{1}\operatorname{sgn}(P_t-\lambda)\dd{\mu}=2\int_{P_t>\lambda}\dd{\mu}-1\,,
\label{eq:threshold}
\end{equation}
where the second equality holds since $\int\dd{\mu}=1$. Each $-1<c_s<1$ selects a threshold, and the corresponding profile $q=\operatorname{sgn}(P_t-\lambda)$ gives the extremum value as well as fixing all Taylor coefficients. At $c_s=\pm1$ the profile is constant
and both sides of \eqref{eq:fixedcs} vanish.


\subsection{Exact bounds for $  c_{s} $ and $  a_{n} $}
\label{sec:paper-parametric-pairs}
We now apply \eqref{eq:fixedcs} to a single coefficient: we fix $c_s$, let every coefficient
other than $a_n$ vary, and construct the allowed region in the $(c_s,a_n)$ plane for $n\ge2$. Denoting by $B_n(c_s)$ the largest allowed $a_n$, the
maximizing profile and the threshold condition \eqref{eq:threshold} are
\begin{equation}
B_n(c_s(\lambda))=2\int_{U_{n-1}>\lambda}U_{n-1}\dd\mu\,,
\label{eq:Bn}
\end{equation}
where we have used $\int U_{n-1}\dd\mu=0$ for $n\geq2$, and the threshold is fixed, as in \eqref{eq:threshold}, with $P_t=U_{n-1}$, and the corresponding profiled is $q=\operatorname{sgn}(U_{n-1}-\lambda)$. Flipping the sign of the maximizing profile for $-c_s$ supplies the minimizing profile for $c_s$, so
\begin{equation}
-B_n(-c_s)\leq a_n\leq B_n(c_s)\,,\qquad n\geq2\,.
 \label{eq:paper-general-pair-bound}
\end{equation}
For even $n$, the polynomial $U_{n-1}$ is odd, so the reflection $q(x)\mapsto q(-x)$ reverses $a_n$ without changing $c_s$. Then $B_n(-c_s)=B_n(c_s)$, and the region is symmetric under either sign reversal. For odd $n$, only the simultaneous reversal of $c_s$ and $a_n$ is a symmetry. To evaluate \eqref{eq:Bn},  let $x_1<\dots<x_J$ be the solutions of
$U_{n-1}(x)=\lambda$ in $(-1,1)$. Using the cumulative functions defined in \eqref{eq:paper-linear-support-antiderivative} we obtain
\begin{tcolorbox}[colframe=white,arc=0pt,colback=greyish2]
\vspace{-0.0cm}
\begin{equation}
c_s=1-2\sum_{j=1}^{J}(-1)^{J-j}K_1(x_j)\,,\qquad
B_n(c_s)=-2\sum_{j=1}^{J}(-1)^{J-j}K_n(x_j)\,,
\label{eq:parametric}
\end{equation}
\end{tcolorbox}
\noindent which describes the boundary parametrically in $\lambda$. We now consider a few special cases.

\begin{figure}[tb]
 \centering
 \includegraphics[width=\textwidth]{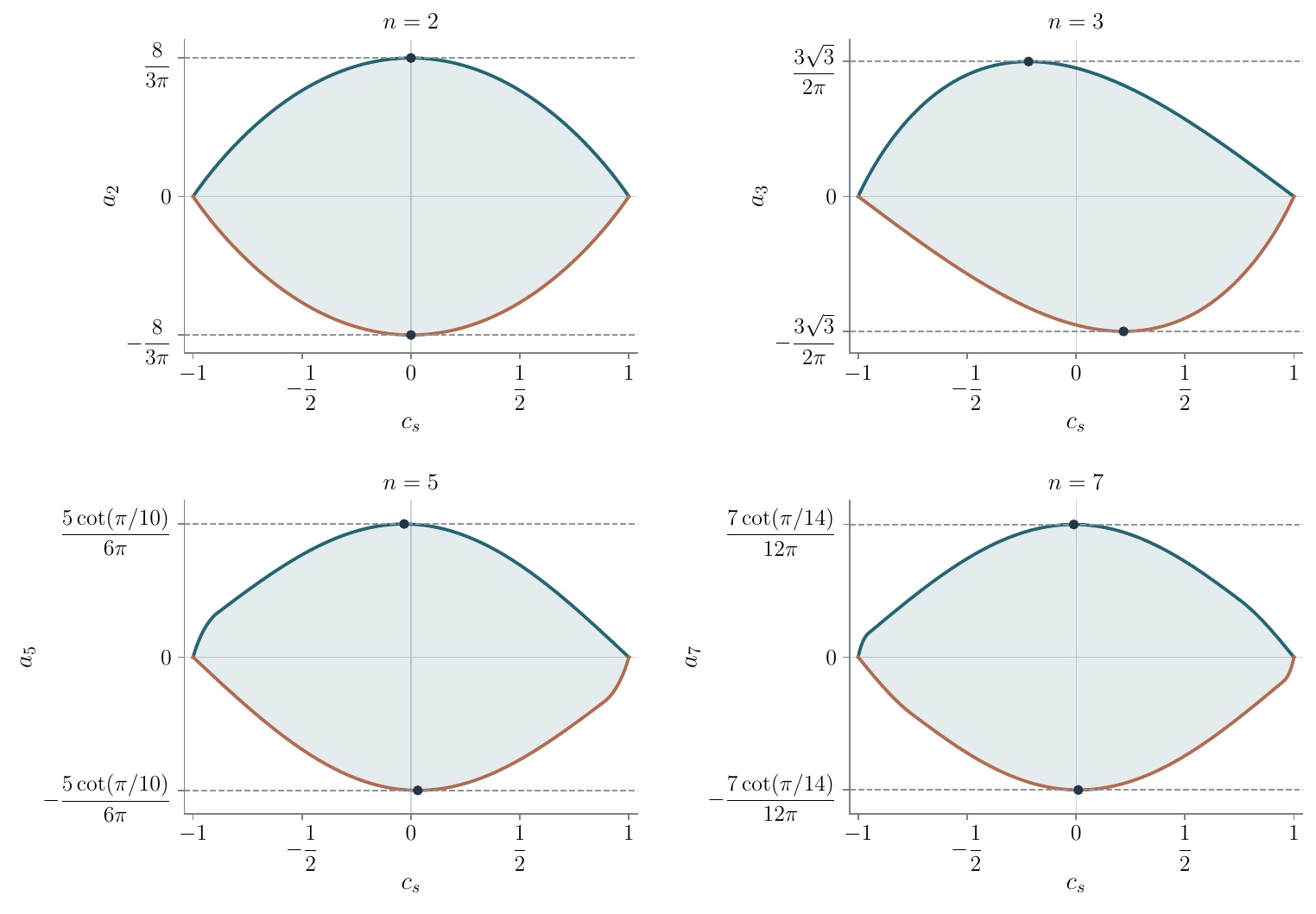}
 \caption{Exact $(c_s,a_n)$ regions for $n=2,3,5,7$, with all other
 Taylor coefficients free. The upper and lower boundaries are shown in
 green and red. The dashed horizontal lines mark the individual
 bounds, whose exact values are given on the vertical axes; the marked
 points show where they are attained. The quadratic region is symmetric
 under either sign reversal. For the odd orders, the symmetry reverses
 $c_s$ and $a_n$ together.}
 \label{fig:paper-pair-regions}
\end{figure}

The simplest case is $(c_{s},a_{2})$. Here $U_1=2x$, with $J=1$, and we define $x_1:=\tau=\lambda/2$. The maximizing
profile is $q=\operatorname{sgn}(x-\tau)$. From \eqref{eq:parametric},
\begin{equation}
c_s=-\frac{2}{\pi}\Big[\arcsin\tau+\tau\sqrt{1-\tau^2}\Big]\,,\qquad
B_2(c_s)=\frac{8}{3\pi}(1-\tau^2)^{3/2}\,,
\end{equation}
where $-1\le\tau\le1$. The space of allowed values for $(c_{s},a_{2})$ is depicted in Figure~\ref{fig:paper-pair-regions}.



For $(c_{s},a_{3})$ we have $U_2=4x^2-1$ and $J=2$. The two solutions of $U_2=\lambda$ are defined $x_{1,2}=\mp b$ with $0\le b\le1$. The maximizing profile is $q=\operatorname{sgn}(|x|-b)$, and we obtain
\begin{equation}
c_s=1-\frac{4}{\pi}\Big[\arcsin b+b\sqrt{1-b^2}\Big]\,,\qquad
B_3(c_s)=\frac{8}{\pi}\,b\,(1-b^2)^{3/2}\,.
\end{equation}
As $b$ decreases from one to zero, $c_s$ increases from $-1$ to $1$, while $a_3$ grows up to
$b=1/2$, and then decreases. The maximum is attained at $c_s=1/3-\sqrt3/\pi$.
Figure~\ref{fig:paper-pair-regions} shows the allowed region, together with those for $n=5,7$ which can be similarly obtained. 


\subsection{Exact bounds for $(c_{s},a_{2},a_{3})$}
\label{sec:paper-three-coefficients}

\begin{figure}[tb]
\centering
\includegraphics[width=\textwidth]{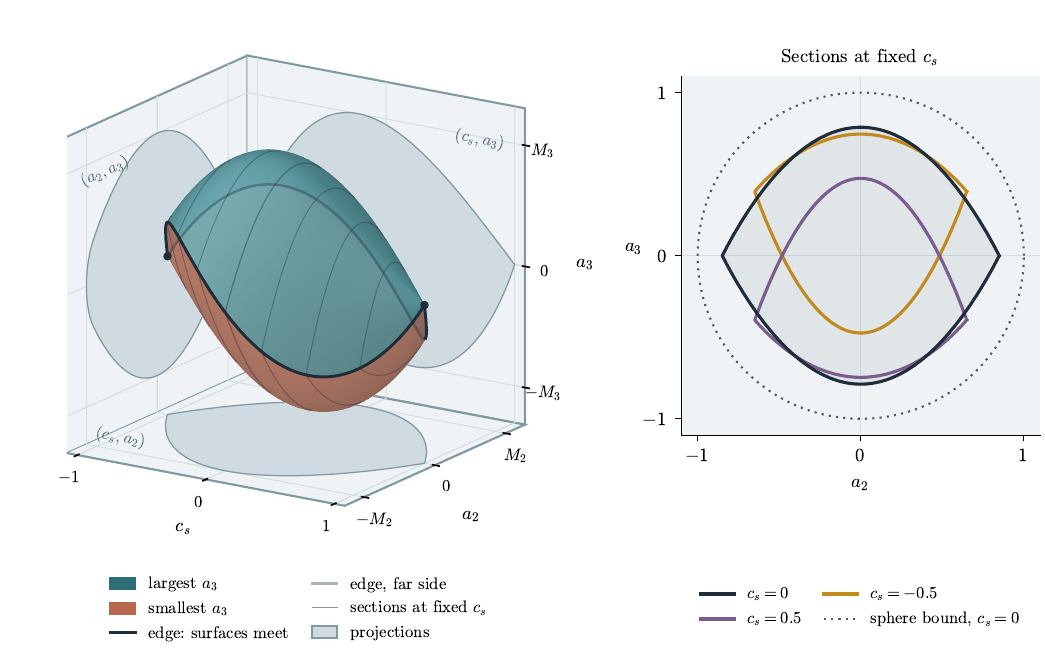}
\caption{The exact $(c_s,a_2,a_3)$ region. Left: the lower and upper surfaces from extremising profiles; the enclosed volume is allowed. Right: sections at fixed $c_s$, compared with the sphere bound at $c_s=0$. The restricted sphere is $c_s^2+a_2^2+a_3^2\le1$. Note that the equality is possible only at $(\pm1,0,0)$, with all the higher coefficients vanish. The other boundary profiles still saturate the \emph{full} sphere bound, with their higher coefficients carrying the missing norm.}
\label{fig:paper-three-coefficients}
\end{figure}

To determine the allowed region for the triple $(c_s,a_2,a_3)$, following section \ref{sec:fixing}, we fix an allowed pair $(c_s,a_2)$ and extremize $a_3$. The shifted polynomial is $Q_\lambda=U_2-\lambda_2U_1-\lambda_1$. The two roots of $Q_\lambda$ within the interval $[-1,1]$, are denoted by $\ell<r$. The the maximum of $a_3$ corresponds to the profile
\begin{equation}
q_{\max}(x)=
  \begin{cases}
    -1, & \ell<x<r,\\
    +1, & \text{otherwise}\,.
\end{cases}
\label{eq:qmax}
\end{equation}
As anticipated, the two Lagrange multipliers can be traded for the two switches $\ell$ and $r$. Evaluating
$c_s=\int q_{\max}\dd\mu$ and $a_2=\int q_{\max}U_1\dd\mu$ with the cumulative functions of \eqref{eq:paper-linear-support-antiderivative}, they read
\begin{equation}
c_s=1-2[K_1(r)-K_1(\ell)]\,,
  \quad
a_2=-2\big[K_2(r)-K_2(\ell)\big]=\frac{8}{3\pi}\Big[(1-r^2)^{3/2}-(1-\ell^2)^{3/2}\Big]\,,
\label{eq:lu}
\end{equation}
with $-1\le\ell<r\le1$. For $-1<c_s<1$ and every $a_2$ in the allowed region, a solution exists and is unique. Evaluating $a_3=\int q_{\max}\,U_2\,d\mu$ with $K_3$ from \eqref{eq:paper-linear-support-antiderivative}
gives the maximum for $a_3$
\begin{equation}
B_3(c_s,a_2)=-2\big[K_3(r)-K_3(\ell)\big]=\frac4\pi\Big[r(1-r^2)^{3/2}-\ell(1-\ell^2)^{3/2}\Big]\,,
\label{eq:a3max}
\end{equation}
where $\ell$ and $r$ are the functions of $(c_s,a_2)$ defined by \eqref{eq:lu}. As argued before, the minimum is given by $-B_3(-c_s,-a_2)$ corresponding to the case where we flip the sign of the extremal profile. Then the allowed region is 
\begin{equation}
-B_3(-c_s,-a_2)\leq a_3\leq B_3(c_s,a_2)\,.
\end{equation}
Together with \eqref{eq:lu}, this is an exact nonlinear description of the allowed region. At $c_s=\pm1$, the triple is simply $(\pm1,0,0)$. Figure~\ref{fig:paper-three-coefficients} shows the resulting volume, together with a comparison with the sphere bound.

The same pattern continues at higher orders. Fixing $c_s,a_2,\dots,a_{N-1}$ and maximizing $a_{N}$ gives a shifted polynomial of degree $N-1$. The extremal profiles have at most $N-1$ switches, and the moment constraints become $N-1$ equations for the switching points.

\section{The convergence radius}
\label{sec:convergence-radius}

In the previous sections, we obtained bounds on the dimensionless coefficients $a_n=\alpha_nR^{n-1}$, normalizing with $R$, which is a scale over which the dispersion branch $\omega(k)$ remains analytic and satisfies the causality condition. Written in terms of the physical parameters, the bounds look like
\begin{equation}
\big(c_s,\,\alpha_2R,\,\cdots,\,\alpha_{N}R^{N-1}\big)\in\mathcal{A}_N\,,
\label{pointAN}
\end{equation}
where the convex space $\mathcal{A}_N$ is fully defined by the infinitely many linear bounds we derived in Section \ref{sec:paper-linear-combinations}. It is then clear that, for fixed physical parameters $\alpha_n$, as $R$ varies, the point in \eqref{pointAN} moves along a curve in the $N-$dimensional space. 

For $R\to0$, the point approaches $(c_s,0,\cdots,0)$ which belongs to $\mathcal{A}_N$ if $-1\leq c_s\leq1$. For $-1< c_s<1$, the point is strictly within the interior of $\mathcal{A}_N$. As a result, for \emph{any} values chosen for $\alpha_2,\cdots,\alpha_N$, we can find small enough $R$ such that the point remains within the allowed neighborhood of $(c_s,0,\cdots,0)$. The endpoints $c_s=\pm1$ are special: sitting at the corner, any scaling with $R$ moves it outside of the allowed space, unless all the other coefficients vanish. In particular, as pointed out in \cite{Creminelli:2026dispersion}, this excludes the possibility of obtaining bounds on combinations of the Taylor coefficients in which the scale $R$ cancels out. Take, for instance, the following combination of rescaled parameters, chosen such that $R$ cancels out,
\begin{equation}
\frac{a_3}{a_2^2}=\frac{\alpha_3}{\alpha_2^2}\,,\quad \frac{a_2a_4}{a_3^2}=\frac{\alpha_2\alpha_4}{\alpha_3^2}\,,\quad\text{etc.}
\end{equation}
There cannot possibly be a bound on any of these combinations since they can be re-written in terms of $\alpha_n$ and we just argue all values for these are allowed for some $R$.\footnote{It is useful to compare with the positivity bounds in the context of scattering amplitudes \cite{Bellazzini:2020cot}. There the low-energy coefficients, $g_n=\int_{M^2}^{\infty}\mathrm{d}s\,\rho(s)/s^{n+1}$, are moments  of a spectral density $\rho(s)\geq0$, which unitarity makes nonnegative. The allowed set is a convex cone, described by homogeneous inequalities, including combinations such as $g_n\,g_{n+2}\ge g_{n+1}^2$ in which $M$ cancels. The difference lies in the scale-invariant theory to which all others flow when the scale is sent to zero. For a gapped amplitude this is necessarily the free theory, $\rho=0$, that saturates positivity, and therefore every deformation is constrained. Here there is family of scale-invariant theories,
$\omega=c_sk$ with $-1\le c_s\le1$. They saturate the causal inequality only at the endpoints $c_s=\pm1$, where in fact a scale-independent bound exists: $\alpha_n=0$ for all
$n\ge2$. For $|c_s|<1$ the causal inequality holds with a finite margin,
$|\!\operatorname{Im}\omega|=|c_s|\,|\!\operatorname{Im}k|$. On a small disk the higher-order terms are small compared to this margin, and no scale-independent constraint on them can arise.} 

In the opposite direction, increasing $R$ gives stronger
constraints on the physical coefficients, as is manifest, for example in $|\alpha_n|\le M_n/R^{n-1}$. The bounds derived above use a disk on which the branch is analytic
and obeys the causal inequality. Therefore, $R$ need not be the actual convergence radius $R_{\mathrm{conv}}$ of the Taylor series. To obtain the strongest bounds one would like to push $R$ all the way up to $R_{\mathrm{conv}}$, which in many examples is set by a branch point where the mode collides with another one at complex momentum
\cite{Withers:2018srf,Grozdanov:2019kge,Grozdanov:2019uhi,Heller:2020hnq}. This includes the possibility of colliding with another member of the family for the same branch (see Section~\ref{sec:paper-branch-symmetries}), as pointed out by \cite{Creminelli:2026dispersion}. We remind that setting $R=R_{\mathrm{conv}}$ assumes that the branch remains a pole of $G_R$ on the physical sheet throughout its disk of convergence; if it crosses a branch cut of $G_R$ at a smaller radius, the causal inequality can only be imposed
up to that radius \cite{Gavassino:2026acausal,Brants:2026causalUV,Creminelli:2026dispersion}.

We now explain when $R$ and $R_{\mathrm{conv}}$ coincide and how knowledge of physical coefficients can constrain the latter. A main conception point, already advocated in \cite{Creminelli:2026dispersion}, is that our bounds should be interpreted as limiting the admissible values of $R_{\mathrm{conv}}$, given that a subset of physical parameters $\alpha_n$ have been measured. 


\subsection{Exact results}
\label{sec:paper-radius-consistency}
Our definition of $R$ is that the dispersion relation is analytic for $|k|<R$. Then the inequality
$R_{\mathrm{conv}}\geq R$ follows by definition, i.e. it is already part of our hypothesis. Conversely, starting from a full profile $-1\leq q\leq1$, the reconstruction in
\eqref{eq:paper-profile-reconstruction} establishes analyticity and
causality throughout this disk. The same conclusion about convergence
follows directly from the coefficients. Since the sphere bound gives
$\sum_{n\geq1}a_n^2\leq1$, the normalized Taylor series converges
absolutely for $|k|<R$. More precisely, Cauchy--Hadamard\footnote{The Cauchy--Hadamard formula states that the power series
$\sum_{n=0}^{\infty} a_n z^n$ has radius of convergence $R$ given by
$R^{-1}=\limsup_{n\to\infty}|a_n|^{1/n}$. It converges absolutely for
$|z|<R$ and diverges for $|z|>R$. Here the limit superior of a sequence $(x_n)$ is defined by $\limsup_{n\to\infty} x_n \equiv \lim_{N\to\infty}\sup_{n\ge N} x_n$.} gives
\begin{equation}
R_{\mathrm{conv}}=
 \frac{R}{\limsup_{n\to\infty}|a_n|^{1/n}}\geq R\,.
 \label{eq:paper-radius-consistency}
\end{equation}
A zero denominator means an infinite radius. This argument uses a profile $q$, or equivalently a full Taylor series, that satisfies our bounds.


\paragraph{The convergence radius of extremizing profiles.}
\label{sec:paper-radius-extremizers}
It turns out that when any of our bounds are saturated, the full dispersion relation is fixed \textit{and} $R=R_{\mathrm{conv}}$. The $a_2$ extremizer \eqref{eq:paper-quadratic-extremal-dispersion} provides a concrete example. Its arctangent
has branch points at $k=\pm iR$, where its prefactor is nonzero.
Equivalently, the coefficients in
\eqref{eq:paper-quadratic-extremal-coefficients} decay as $1/j$, so the
limsup in \eqref{eq:paper-radius-consistency} equals one. Then, specifying
its positive physical coefficient fixes the scale,
\begin{equation}
 R_{\mathrm{conv}}=R=\frac{8}{3\pi\alpha_2},\qquad \alpha_2>0.
 \label{eq:paper-quadratic-radius}
\end{equation}
A similar equality holds for every branch extremizing an individual
coefficient of order $n\geq2$. 
There is an intuitive way to understand why extremizing profiles have $R =R_{\mathrm{conv}} $. Extremizing profiles have discontinuities when their value flips from -1 to 1 or vice versa. Therefore, also the dispersion relation on the disk $|k|=R$ needs to have corresponding discontinuities. It therefore cannot be analytical and hence $R=R_{\mathrm{conv}}$. The argument also applies
to the fixed-$c_s$ extrema of Section~\ref{sec:paper-parametric-pairs}
for $-1<c_s<1$, as well as to extrema of linear combinations of coefficients since in all of these cases, the profiles have discontinuities. By the same token, every causal branch saturating the sphere
bound has $R_{\mathrm{true}}=R$, with the only exception of $\omega=\pm k$, whose profile has no discontinuity, $q=\pm 1$, and which is analytic on the whole plane.


\subsection{Upper bounds}
\label{sec:paper-radius-coefficients}

We now assume that the branch obeys the causal inequality throughout
its true convergence disk $|k|
<R_\text{conv}$. Applying the individual bound
\eqref{eq:paper-individual-bound} gives\footnote{No regularity on the limiting circle is required. To see this, impose the bounds at every radius $r<R_{\mathrm{conv}}$ and take $r\uparrow R_{\mathrm{conv}}$.}
\begin{equation}
 |a_n|=|\alpha_n|R_{\mathrm{conv}}^{n-1}\leq M_n,
 \implies \displaystyle R_{\mathrm{conv}}\leq
 \left(\frac{M_n}{|\alpha_n|}\right)^{1/(n-1)},
 \qquad n\geq2,\quad\alpha_n\ne0.
 \label{eq:paper-radius-single-bound}
\end{equation}
The bound is saturated by the corresponding extremizer, such as for example \eqref{eq:paper-quadratic-radius}. Note that dimensionless coefficient $c_s$ by itself supplies no bound on $R_{\mathrm{conv}}$. If several coefficients there are many different ways to derive an upper bound on $R_{\mathrm{conv}}$ using our results. Instead of always establishing the optimal one, we comment below on a few possibilities. 

Say we measure $\alpha_n$ for $n\in S$. Then we can take the minimum of their individual
upper bounds as the best upper bound $R_\text{ind}$
\begin{align}
    R_{\text{ind}}:=\min_{n\in S}\left( \frac{M_n}{|\alpha_n|}\right)^{1/(n-1)}\,.
\end{align}
However this is in general not optimal: joint constraints can strengthen the upper bound on $R_\text{conv}$. For example, the sphere bound implies
\begin{equation}
 c_s^2+\sum_{\substack{n\geq2\\n\in S}}
 \alpha_n^2R_{\mathrm{conv}}^{2n-2}\leq1.
 \label{eq:paper-radius-finite-sphere}
\end{equation}
hence determining a best upper bound $R_{\mathrm{conv}}\leq R_\text{sph}$ from the sphere bound and the knowledge of $\alpha_n$ with $n\in S$. Depending on the values of $\alpha_n$, the sphere bound may provide a stronger or weaker upper bound on $R_{\mathrm{true}}$ than $R_{\mathrm{ind}}$. The maximum improvement is quantified by $R_\text{sph}/R_\text{ind}$, which is minimized when all the individual bounds from $\alpha_n$ give the same $R_\text{ind}$. In this case the improvement is the first positive real root of 
\[
\sum_{\substack{n\geq2\\n\in S}} M_n^2 x^{2(n-1)}=1-c_s^2 .
\]
For $c_s=0$ the improvement is approximately between 10\% and 30\%, but becomes substantial for $c_s \to 1$. If the value of $c_s$ is known together with $\alpha_n$ with $n\in S$ then one would use some of the joint bounds we have derived in the Section \ref{sec:nonlinear-bounds}. For example, the exact pair and triple bounds of Section~\ref{sec:nonlinear-bounds} can be applied with $a_n=\alpha_nR_{\mathrm{true}}^{n-1}$.

Finally, if all coefficients are known exactly, the individual bounds alone
recover the exact radius of convergence. Indeed using \eqref{eq:paper-radius-consistency} as well as the limit $M_n\to8/\pi^2$ for $n\to \infty$ gives
\begin{equation}
 \liminf_{n\to\infty}
 \left(\frac{M_n}{|\alpha_n|}\right)^{1/(n-1)}
 =\frac{1}{\limsup_{n\to\infty}|\alpha_n|^{1/n}}
 =R_{\mathrm{conv}}.
 \label{eq:paper-radius-recovery-proof}
\end{equation}
We stress that for all these upper bounds it is essential that the dispersion relation obeys causality everywhere inside the radius of convergence.

\section{Causality outside the convergence disk}
\label{sec:global-causality}

The bounds derived so far use microcausality only inside the disk $|k|<R$. In
this section, we ask what is gained by going beyond it. First, we look at two examples that show that causality on the disk can be much weaker compared to all momenta. We then show that the branches saturating our bounds violate causality immediately outside the disk, so that the inequalities become strict. These strict bounds are nevertheless sharp, since they are approached arbitrarily closely by poles of globally causal Green functions.

\subsection{A massive scalar}

Consider a free scalar with dispersion relation
$\omega(k)=\sqrt{m^2+v^2k^2}$, with $m>0$ and $v^2>0$. The expectation is that causality must require $|v|\leq 1$. On the other hand, it is straightforward to check that restricting to the convergence disk gives a much weaker bound. The branch points at $k=\pm i m/|v|$ set the convergence radius $R_{\mathrm{conv}}=m/|v|$. Expanding around $k=0$ gives the dimensionless coefficients\footnote{For arbitrary $\alpha$ and integer $n\geq1$, we define $\binom{\alpha}{n}=\alpha(\alpha-1)\cdots(\alpha-n+1)/n!$.}
\begin{equation}
a_{2n}=\binom{1/2}{n}\left(\frac{R}{R_{\mathrm{conv}}}\right)^{2n-1}|v|\,,
\qquad n\geq1\,,
\end{equation}
with all the odd coefficients vanishing. For the best bound we set $R=R_{\mathrm{conv}}$. If we consider bounds on the individual coefficients, i.e. \eqref{eq:paper-individual-bound}, the strongest comes from $a_2$ which gives $|v|\leq16/(3\pi)$.

We can improve this bound by considering all possible linear combinations of Section~\ref{sec:paper-linear-combinations}. Equivalently, we can evaluate the profile $q(x)$ corresponding to the dispersion relation and require $|q|\leq1$. 
Since $\omega(R_{\rm conv}e^{i\theta})=m\sqrt{1+e^{2i\theta}}$, the profile
in \eqref{eq:paper-massive-profile-example} acquires a factor of $|v|$,
\begin{equation}
 q(x)=|v|\operatorname{sgn}(x)\sqrt{\frac{|x|}{1+|x|}}\,.
\end{equation}
Using $|q(x)|\leq 1$ the strongest bound, from causality within the disk, is $|v|\leq\sqrt{2}$ which is still weaker compared to $|v|\leq1$. Note that even for $|v|=\sqrt{2}$, all bounds on the linear combinations discussed in \eqref{eq:paper-linear-support} are satisfied with strict inequality, since $|q(x)|<1$ for $-1<x<1$. Our bounds allow for values $1<|v|\leq\sqrt{2}$ because a violation of causality happens outside the convergence radius.

It is possible to get the optimal bound by imposing causality \textit{outside} of the convergence disk, since we know the complete form of the dispersion relation. More generally, if we know all the Taylor coefficients $a_n$, we can resum the series and analytically continued outside of the radius of convergence. For the current example, we can follow the
dispersion relation along the real axis to arbitrarily large $k$. For real dispersion relations the group velocity must be bounded by one, so we find
\begin{equation}
 \omega'(k)
 =\frac{v^2 k}{\sqrt{m^2+v^2k^2}}\leq1 .
\end{equation}
As $k\to+\infty$, the group velocity tends to $|v|$, so global
causality requires $|v|\leq1$. One can show that this is indeed optimal, by explicitly checking $\operatorname{Im}\omega(k)\leq|\operatorname{Im}k|$ in the whole complex plane. This simple example highlights that the disk of convergence of $\omega(k)$ is not the same as the regime of validity of the theory.\footnote{See the discussion in \cite{Creminelli:2026dispersion} for the distinction between $\omega$ and $\omega^2$.} Hence bounds arising from causality in the disk are generically not optimal. 


\subsection{The gapless branch of the Drude--Lorentz model}

Here's an even more extreme example. Consider the Drude--Lorentz model of Ref.~\cite{Creminelli:2026dispersion}, with $M,g>0$ and $c_r\geq0$. Its Lagrangian and pole equation are
\begin{align}
 \mathcal L
 &=\frac12\left[\dot\pi^2-(\nabla\pi)^2+\dot h^2
       -c_r^2(\nabla h)^2-M^2h^2\right]-gh\dot\pi \,,
 \nonumber\\
 \det G_R^{-1}
 &=(\omega^2-k^2)(\omega^2-c_r^2k^2-M^2)-g^2\omega^2=0\,.
 \label{eq:drude-poles}
\end{align}
It is useful to define $\Delta^2=M^2+g^2$ and $c_s=M/\Delta$. We mainly focus on the gapless dispersion branch which is $\omega_-(k)=c_s k+O(k^3)$. Microcausality of the full theory gives the bound $c_r\leq 1$. However, perhaps surprisingly, demanding microcausality inside the convergence radius, with $R=R_{\rm conv}$, provides no bound on $c_r>1$ whatsoever. 

For example, at $c_s=1/2$ and $c_r\geq1$, the radius
is $R_{\mathrm{conv}}=\Delta/(2c_r)$ and $a_3=(3/16)(1-1/(4c_r^2))$. At $c_r=2$
this gives $a_3=45/256$. Even as $c_r\to\infty$, the first two
Taylor coefficients approach $a_3=3/16$ and $a_5=-21/256$. All these values are well within the individual bounds \eqref{eq:paper-individual-bound}. In fact, it can be shown that the whole gapless branch\footnote{Imposing causality for the \textit{gapped} branch inside the convergence disk constrains $c_r < 1.59 $, which is better, yet still suboptimal. } obeys causality \textit{inside its convergence disk} for arbitrarily large $c_r$.  

Understanding how $\omega_-$ manages to remain causal in its convergence radius reveals an important distinction. At an unspecified radius $R$ one has
\begin{align}
a_1=c_s,\qquad a_{2n}=0,\qquad
a_3=\frac{(1-c_s^2)(c_r^2-c_s^2)}{2c_s}
       \frac{R^2}{\Delta^2}\,.
\end{align}
So $a_3$ would grow with $c_r$ \textit{at fixed} $R$. However for \(c_r\ge c_s/\sqrt{1-c_s^2}\) we have $R=\Delta c_s/c_r$ so the convergence disk shrinks as $c_r$ grows and $\omega_-$ remains microcausal inside of it. Despite these observations, our bounds are still useful as constraints on $R_\text{conv}$, as stressed in \cite{Creminelli:2026dispersion}. Indeed, a measurement of $\alpha_3$ implies
\[
R_{\mathrm{conv}}\le
\sqrt{\frac{3\sqrt3}{2\pi|\alpha_3|}}\,,
\]
even without any knowledge of the parent theory \eqref{eq:drude-poles}.

As in the previous example, we can do much better if, by virtue of knowing the full Taylor expansion or otherwise, we can impose causality also outside the convergence disk. Notice that, as also stressed in \cite{Creminelli:2026dispersion} this is perfectly appropriate: $G_R$ is meromorphic so branches cannot leave the physical sheet. Moreover the original theory is Gaussian and valid for all kinematics. To find the optimal bound, we follow the same gapless branch along the positive real axis for $c_r>1$. At large $k$ we can compute the group velocity
\begin{equation}
 \omega_-(k)=k-\frac{g^2}{2(c_r^2-1)k}+O(k^{-3})\,,\quad\implies\quad
 \omega_-'(k)=1+\frac{g^2}{2(c_r^2-1)k^2}+O(k^{-4})\,.
\end{equation}
This violates causality for all $c_r>1$. Conversely, a subluminal group velocity requires $c_r\leq1$, which is the optimal bound from global causality on the gapless branch.


\subsection{Excluding saturating profiles}\label{extdisp}

Here we use the idea of imposing microcausality outside of the convergence radius to exclude any profile taking the value one or minus one in an open set. In particular, this excludes the extremizing profiles we have derived in earlier sections. The idea is that whenever $q(x)=1$ on any open interval of $x$, the corresponding dispersion relation violates microcausality right outside the disk. To prove this we use \eqref{eq:paper-profile-reconstruction} to analytically continue in $k=Re^{i\theta}$ outside the radius of convergence and then test microcausality. We have 
\begin{equation}
\omega(k)-k=-k\int_{-1}^1
\frac{1-q(x)}{1-2x(k/R)+(k/R)^2}\dd{\mu(x)}\,.
\end{equation}
We note that if $x=\cos\theta$ lies inside the arc where $q=1$, the numerator vanishes in a neighborhood of that point. The integral therefore continues analytically as written across the arc and outside the disk. We can then take imaginary part, or simply use \eqref{eq:paper-reconstruction-average} and \eqref{tobeadded}, and  re-arrange the terms to write
\begin{equation}
 \frac{\operatorname{Im}\omega(k)}{\operatorname{Im}k}-1
 =\left(\frac{|k|^2}{R^2}-1\right)
 \int_{-1}^1
 \frac{1-q(x)}{|1-2x(k/R)+(k/R)^2|^2}\,\dd{\mu(x)}\,.
\end{equation}
We further assume that the dispersion relation is not linear, i.e. $\omega(k)\neq  k$.\footnote{This in particular implies $R=R_{\mathrm{conv}}$ because if by contradiction $R<R_{\mathrm{conv}}$ then $q(\cos\theta)$ would have been real analytic at $|k|=R$ for $0<\theta<\pi$, so its value $q=1$ on the open arc would fix it to one throughout, contradicting $\omega(k)\neq k$.} Therefore the profile cannot everywhere be $q(x)=1$, which implies that the integral on the right-hand side is strictly positive. Then immediately outside
the upper arc, $|k|>R$ we find the violation $\operatorname{Im}\omega(k)>\operatorname{Im}k$. A similar argument excludes profiles that take $q=-1$ on an open interval.\footnote{An equivalent argument is as follows. Over the saturating interval, $\omega-k$ is real. In that case, Shwarz reflection gives the exact analytic continuation outside $\omega(k)=k-\frac{R^2}{k}+\omega(R^2/k)$. Causality condition is violated outside, since $\operatorname{Im}(\omega(k)-k)=\operatorname{Im}(\Bar{k}-\omega(\Bar{k}))>0$, with $\Bar{k}=R^2/k^*$ is the mirror point which lies inside of the disk where causality condition holds.}

This argument excludes the extremizing profiles we found around \eqref{eq:paper-extremal-dispersion}. In particular, the profiles extremizing $a_n$ for $n\geq2$ equal one
on an open interval. Therefore they violate causality just outside the radius of convergence. This further consideration turns all of our bounds in this paper into strict bounds. For example
\begin{align}
    |a_n|&< M_n \,, & \left|\sum_{n=1}^N t_na_n\right|
&<\sum_{j=0}^{J}\left|K_t(x_{j+1})-K_t(x_j)\right|\,, & \text{etc.}
\end{align}
The natural question now is whether these strict bounds are sharp, i.e.~if one can approach the bound arbitrarily closely, or if they can be further improved by considerations of causality outside the radius of convergence. 


\subsection{Approaching the bounds with globally causal Green's functions}
\label{sec:approaching-causal-bounds}

Here we construct the family of dispersion relations that we can get arbitrarily close to saturating the above strict bounds, which are therefore sharp. First we briefly introduce the general formalism and then present the key element of the construction. Many more details are given in Appendix \ref{app:quadratic-sharpness}.

Consider\footnote{EP thanks Lorenzo Gavassino for pointing out this construction.} the family of rational Green functions\footnote{To impose parity one includes \( D(-\omega,k)D(\omega,-k) \) in the denominator before taking the retarded boundary value. }
\begin{equation}
 D(\omega,k)=\det(\omega I-A-kB)\,,\qquad
 G_R(\omega,k)=\frac{1}{D(\omega,k)D(-\omega,-k) }\,,
 \label{eq:GRfamily}
\end{equation}
with the retarded prescription. Here, \(A\) and \(B\) are real symmetric  $N\times N$ matrices , with the eigenvalues of $B$ in $[-1,1]$. At a zero of $D$ there is a vector $u\neq0$ with $(A+kB)u=\omega u$. Contracting with $u^\dagger$ and taking the imaginary part gives
\begin{equation}
 \operatorname{Im}\omega
 =\operatorname{Im}k\,\frac{u^\dagger Bu}{u^\dagger u}\,,\quad\implies\quad
|\operatorname{Im}\omega|\leq|\operatorname{Im}k|\,,
\label{ABeigen}
\end{equation}
where we used $u^\dagger A u$ and $u^\dagger B u$ are real and eigenvalues of B lie in $[-1,1]$. This holds at every complex \(k\), and likewise for the poles of
\(1/D(\omega,-k)\). Hence every pole of $G_R$ obeys the causal inequality in the
whole complex $k$ plane, and for real $k$ all poles are real. Intuitively, \eqref{eq:GRfamily} describes $N$
modes that propagate with velocities given by the eigenvalues of $B$ and are
mixed by the constant matrix $A$. These can be realized by $i\partial_t\psi=(A-iB\partial_x)\psi$. Note that since $G_R$ is rational, its poles can be continued in $k$ without ever leaving the physical sheet. 

As an example, now we use this formalism to build a sequence of dispersion relations saturating \(a_2\). We denote by \(R_{\text{conv}}\) be the convergence radius of the extremizing branch which maximizes $a_2$. We want to construct a sequence of poles of \eqref{eq:GRfamily}, denoted by $\omega_N(k)$, whose quadratic coefficient approaches this bound. The idea, which is explained more pedagogically in Appendix \ref{app:quadratic-sharpness}, is to create a sequence of discrete sampling of \eqref{eq:paper-profile-reconstruction} that becomes finer and finer. To this end we sample the extremal profile on the upper semicircle \(|k|=R_{\text{conv}}\) at the angles \(\theta_j=j\pi/(N+1)\), \(j=1,\ldots,N\). The matrix \(B\) is chosen to be diagonal with diagonal elements given by the sampled values \(\operatorname{sgn}(\cos\theta_j)\). The matrix \(A\) is more complicated and given in Appendix~\ref{app:quadratic-sharpness}. Solving for the ensuing dispersion relation in perturbation theory near $k=0$ we find
\begin{equation}
 \alpha_{2,N}=\frac{4}{R_{\text{conv}}(N+1)}\sum_{j=1}^N
 \sin^2\theta_j\,|\cos\theta_j|\,.\label{7p11}
\end{equation}
Since the angular spacing is \(\Delta\theta=\pi/(N+1)\), for large $N$ this sum approaches the following elementary integral
\begin{align}\label{eq:limit}
    \lim_{N\to \infty}\alpha_{2,N}=\frac{4}{\pi R_{\text{conv}}}\int_0^\pi
 \sin^2\theta\,|\cos\theta|\,\dd{\theta}
 =\frac{8}{3\pi R_{\text{conv}}}=\frac{M_2}{R_{\text{conv}}}\,.
\end{align}
To conclude the proof that our strict bound on $\alpha_2$ is sharp, Appendix \ref{app:quadratic-sharpness} gives \(A/R_{\text{conv}}\) explicitly and shows that the true convergence radius \(R_N\) of this $\omega_N$ tends to \(R_{\text{conv}}\) for large $N$.
Thus its dimensionless coefficient \(a_{2,N}=\alpha_{2,N}R_N\)
approaches \(M_2\).

A similar construction applies for all other Taylor coefficients. Sampling \(q=\pm\operatorname{sgn}U_{n-1}\) in the same way gives a sequence of dispersion relations approaching either endpoint for any fixed \(n\geq2\). For brevity we don't present here this construction. We conclude that the bound \(|a_n|<M_n\) is sharp for dispersion relations that are \textit{globally} microcausal: no smaller constant can replace \(M_n\).

A closing remark is in order. The branches $\omega_N$ are globally causal for all $N$, while their limit, the extremal branch, violates causality just outside the disk, as argued in the previous subsection. However, there is no contradiction because $\omega_N\to\omega$ only inside the disk, where the extremal profile does obey microcausality. Each $\omega_N$ has branch points
where it collides with other poles of \eqref{eq:GRfamily}. These lie all around the circle $|k|=R_{\rm conv}$, at a distance of order $R_{\rm conv}\log N/N$ from it. As $N$ grows they accumulate on the whole circle, including the arcs across which the limiting branch can be analytically continued. The set of globally causal branches is therefore not closed. This is why the bounds are strict and yet sharp.

\section{Conclusion}
\label{sec:conclusion}

We have studied the constraints that relativistic causality imposes on a
non-dissipative dispersion branch $\omega(k)$ that is analytic around
$k=0$. Microcausality requires
$|\!\operatorname{Im}\omega|\le|\!\operatorname{Im}k|$ on the disk
$|k|<R$ where the branch is analytic and remains a pole of the retarded
Green function on the physical sheet. The main observation is that this
condition is equivalent to $-1\le q\le1$ where we introduced
the profile $q=\operatorname{Im}\omega/\operatorname{Im}k$ on the circle $|k|=R$. The dimensionless Taylor coefficients $a_n=\alpha_nR^{n-1}$ are the Chebyshev moments of $q$, and
every bounded profile reconstructs an analytic and causal branch. The
problem of causality bounds thus becomes a Markov moment problem.

We obtained sharp bounds on individual
coefficients together with the corresponding extremal dispersion relations in closed form. Also obtained the complete set of linear inequalities that are necessary and
sufficient for a set of coefficients to be allowed. When some coefficients are fixed, the same principle applies after introducing one Lagrange multiplier for each of them. In this way we obtained the exact regions for the pairs $(c_s,a_n)$ and for the triple $(c_s,a_2,a_3)$. Finally, all
coefficients obey the sphere bound $\sum_na_n^2\le1$, which expresses causality on average over the circle. In all cases, saturating a bound fixes the entire dispersion relation.

We would like to make three important comments on our results. First, apart from $|c_s|\le1$, all bounds relate the coefficients to the radius $R$, and no combination of finitely many coefficients is constrained independently of it. The bounds are best read as upper limits on the scale at which the derivative expansion of the mode must break down: if the branch stays on the physical sheet throughout its disk of convergence, each measured coefficient gives $R_{\rm conv}\le(M_n/|\alpha_n|)^{1/(n-1)}$, and joint bounds are stronger. Second, the bounds are sharp for the problem on the disk, but they do not exhaust the constraints of causality. The examples of Section~\ref{sec:global-causality} show that following a branch beyond its disk of convergence can give stronger conditions, and the extremal dispersion we have derived violate causality right outside the convergence disk. A branch that remains a pole of a causal Green function beyond the disk therefore obeys our bounds with a strict inequality.

Several directions are left for future work. The most immediate one is to use information beyond the disk. This requires assumptions on where the continued branch can be singular, and it would be interesting to understand which of them are generic enough to yield universal bounds. As pointed out in \cite{Creminelli:2026dispersion}, instead of $\omega(k)$, one can study implications of causality on $\omega^2(k)$. The bounds obtained on the latter are naturally interpreted as a bound on the scale of new physics. It would be interesting to obtain analytic results in that case. Another direction is adding dissipation, i.e.~complex Taylor coefficients  where the profile is no longer bounded from below.

\paragraph{Acknowledgement} It is a pleasure to thank Trevor Cheung, Paolo Creminelli, Carlos Duaso Pueyo, Lorenzo Gavassino, Harry Goodhew, Michael Mancini, Warin P. McBlain, and Benjamin Withers for useful discussion. OpenAI’s GPT-6 Astra and Anthropic's Opus-5.5 played a key role in the derivation of our results. All results were critically examined and verified by the authors. This work has been supported by STFC consolidated
grant ST/X001113/1, ST/T000694/1, ST/X000664/1 and ST/Y509127/1.

\appendix

\section{Approaching microcausality bounds}
\label{app:quadratic-sharpness}

In this appendix we give the matrices \(A\) and \(B\) used in Sec.~\ref{sec:approaching-causal-bounds}. To approximate the dispersion branch, we discretize \eqref{eq:paper-profile-reconstruction} assuming some given profile. We will keep $R$ generic for now but will eventually set it to $R_{\rm conv}$. We choose the following set of discrete points on the upper semicircle where we sample the integrand
\begin{equation}
  \theta_j=\frac{j\pi}{N+1}\,,\qquad x_j=\cos\theta_j\,,\qquad
  b_j=q(x_j)\,,\qquad p_j=\frac{2}{N+1}\sin^2\theta_j\,,
  \label{eq:nodes}
\end{equation}
for $1\le j\le N$. The order of approximation is controlled by $N$, which we will take the infinity limit, and sets the dimension of the matrices $A$ and $B$. The sampled values of the profile are $b_j\in[-1,1]$, while $p_j$ represents the discretization of the measure $d\mu$ and satisfying $\sum_jp_j=1$. As we shall see later, rather than specifying the dispersion relation by $\omega(k)$, it is simpler to write $k(z)$ and $\omega(z)$ parameterized with some parameter $|z|<1$. In the continuum limit, we want $k(z)=Rz$ but it can be more complicated at order $N$ approximation. From the denominator of \eqref{eq:paper-profile-reconstruction} we have: $1-2x(k/R)+(k/R)^2\to 1-2xz+z^2$. As a result, in the discrete approximation we define
\begin{equation}
d_j(z)=1-2x_jz+z^2\,.   
\end{equation}
Then the discretized versions of \eqref{eq:paper-profile-reconstruction} becomes
\begin{equation}
\omega_N(z)=Rz\sum_j\frac{p_j\,b_j}{d_j(z)}\,,\qquad k_N(z)=Rz\sum_j\frac{p_j}{d_j(z)}= Rz\,\frac{1-z^{2N}}{1-z^{2N+2}}\,,
\label{eq:target}
\end{equation}
where we have called the $N-$th order approximate functions $\omega_N(z)$ and $k_N(z)$. The expression for the latter is obtained simply by setting $q(x)=1$ in the continuum and then discretize, i.e. $b_j=1$. The second equality for $k_N(z)$ follows from expanding $1/d_j(z)=\sum_{m}U_m(x_j)\,z^m$ and using the discrete
orthogonality of the Chebyshev polynomials, $\sum_jp_jU_m(x_j)U_n(x_j)=\delta_{mn}$, for $0\leq m,n\leq N-1$. Taking the imaginary part we have 
\begin{equation}
\operatorname{Im}\omega_N
=\operatorname{Im}k_N\frac{\sum_j b_j\;p_j/|d_j(z)|^2}{\sum_j p_j/|d_j(z)|^2}\,,
\label{imoimk}
\end{equation}
which is discrete version of \eqref{eq:paper-reconstruction-average}. 

Next, we construct, at each order $N$, a Green function of the globally causal family \eqref{eq:GRfamily} that has dispersion branch of \eqref{eq:target}. Then, almost by construction, $\omega_N$, for $N\to\infty$, approaches the actual dispersion relation we want to approximate. We need to specify appropriate $A$ and $B$ matrices. 

We begin by taking $B$ to be a diagonal matrix of size $N$. This is no loss of generality as we can always diagonalize a real symmetric matrix. The natural candidate for diagonal entries are $b_j$'s, i.e. $B=\operatorname{diag}(b_1,\dots,b_N)$, which as required are between $-1$ and $1$. The intuition is that we need $N$ modes to approximate the dispersion branch whose velocities are given by sampling of the profile $q(x)$. From the eigenvalue equation \eqref{ABeigen} we have
\begin{equation}
  \frac{\operatorname{Im}\omega_N}{\operatorname{Im}k_N}
  =\frac{u^\dagger Bu}{u^\dagger u}
  =\frac{\sum_j b_j\,|u_j|^2}{\sum_j|u_j|^2}\,.
  \label{eq:velocityaverage}
\end{equation}
Comparison with \eqref{imoimk} gives that the corresponding eigenvector must be 
\begin{equation}
u_j(z)=\frac{\sqrt{p_j}}{d_j(z)}\,.
\label{eq:Bandv}
\end{equation}
We need to find a consistent symmetric matrix $A$, such that $u(z)$ is an eigenvector $(A+Bk_N)u=\omega_N u$ for all
$|z|<1$. The eigenvalue equation can be written
\begin{equation}
\sum_jA_{ij}\sqrt{p_j}
-2z\sum_{j\neq i}\frac{A_{ij}\sqrt{p_j}\,(x_i-x_j)}{d_j(z)}
=\Big(\omega_N(z)-k_N(z)\,b_i\Big)\sqrt{p_i}\,.
  \label{eq:split}
\end{equation}
To obtain \eqref{eq:split} we have multiplied the equation by $d_i$ and used $d_i/d_j=1-2z\,(x_i-x_j)/d_j$. By using \eqref{eq:target} we have $\omega_N(z)-k_N(z)\,b_i=Rz\sum_{j\neq i}p_j(b_j-b_i)/d_j$. Then it is straightforward to obtain
\begin{equation}
  A_{ij}=\frac{R_{\mathrm{conv}}}{2}\sqrt{p_i p_j}\,,
         \frac{b_i-b_j}{x_i-x_j}\quad(i\ne j),\qquad
  A_{ii}=-\frac{R_{\mathrm{conv}}}{2}
         \sum_{j\ne i}p_j\,\frac{b_i-b_j}{x_i-x_j}\,,
  \label{eq:quadratic-matrices}
\end{equation}
where we have finally set $R=R_{\rm conv}$. This concludes the construction. 

In order to obtain the Taylor coefficients one has to invert \eqref{eq:target} and obtain $\omega_N(k_N)$. It is useful first to see an example for $N=2$. From \eqref{eq:nodes}, $x_{1,2}=\pm\frac{1}{2}$ and $p_{1,2}=\frac{1}{2}$. We choose $b_{1,2}=\pm1$, then we have
\begin{equation}
A=\frac{R_{\rm conv}}{2}\begin{pmatrix}
-1&1\\1&-1
\end{pmatrix}\,,\qquad B=\begin{pmatrix}
1&0\\0&-1
\end{pmatrix}\,.
\label{ABmat}
\end{equation}
The parametrized dispersion relation \eqref{eq:target} is
\begin{equation}
k(z)=\frac{R_{\rm conv}z}{1+z^2+z^4}\,,\qquad\omega(z)=\frac{R_{\rm conv}z^2}{1+z^2+z^4}\,,
\end{equation}
or $\omega(k)=-\frac{R_{\rm conv}}{2}\pm\sqrt{R_{\rm conv}^2/4+k^2}$, which can also be obtained directly from \eqref{ABmat}. For the gapless branch, we obtain the expansion $\omega(k)=k^2/R_{\rm conv}-k^4/R_{\rm conv}^3+\cdots$, corresponding to $\alpha_2=1/R_{\rm conv}$. Since the radius of convergence, at this order, is $R_{\rm conv}/2$, the dimensionless Taylor coefficient is $a_2=1/2<M_2$ consistent with the bound in \eqref{eq:paper-individual-bound}. More generally, to obtain $\omega_N(k_N)$, one has to solve for $z(k_N)$ and plug this into $\omega_N(z)$ which can be done perturbatively for small $z$. Using \eqref{eq:target}, the $n-$th Taylor coefficient is obtained as
\begin{equation}
\alpha_{n,N}\,R_{\rm conv}^{\,n-1}=\sum_{j=1}^Np_j\,b_j\,U_{n-1}(x_j)
\;\xrightarrow[N\to\infty]{}\;\int_{-1}^{1}q(x)\,U_{n-1}(x)\,d\mu(x)\,,
\qquad n\le2N\,.
\label{eq:anN}
\end{equation}
For $n=2$, this is precisely \eqref{7p11}.

To finish the proof we have to show that the radius of convergence of \(\omega_N(k)\) asymptotes $R_\text{conv}$ at large $N$. Since $\omega_N(z)$ is analytic for
$|z|<1$, analyticity in $k$ can be lost only where $z\mapsto k_N(z)$ fails to be invertible. Let \(R_N\) be the convergence radius of $\omega_N$ and fix \(0<r<1\). At \(k=0\), the only solution of \(k_N(z)=0\) inside \(|z|=r\) is \(z=0\). As \(k\) varies, the number of solutions inside the circle, counted with multiplicity, can change
only if a solution crosses the circle. A crossing requires \(k=k_N(z)\) for some \(|z|=r\). Thus, for
\[
  |k|<\min_{|z|=r}|k_N(z)|,
\]
there is still exactly one solution inside the circle, which depends analytically on \(k\). Substituting the solution \(z=z(k)\) of \(k_N(z)=k\) into $\omega_N$ in \eqref{eq:target} gives an analytic dispersion relation throughout this
disk, hence putting a lower bound on $R_N$. Together with the upper bound we derived in the main text, this gives the two sided bound
\[
  \min_{|z|=r}|k_N(z)| \;\le\; R_N
  \;\le\; \frac{M_n}{\alpha_{n,N}}\,.
\]
The lower bound tends to \(rR_{\mathrm{conv}}\), since
\(k_N(z)\to R_{\mathrm{conv}}z\) uniformly on \(|z|=r<1\).
The upper bound tends to \(R_{\mathrm{conv}}\) by
\eqref{eq:anN}. Since \(r\) can be taken
arbitrarily close to one, \(R_N\to R_{\mathrm{conv}}\), and hence
\(a_{n,N}=\alpha_{n,N}R_N\to M_n\).

\phantomsection
\addcontentsline{toc}{section}{References}
\bibliographystyle{references/JHEP}
\bibliography{references/paper.bib}

\providecommand{\href}[2]{#2}\begingroup\raggedright\begin{thebibliography}{10}

\bibitem{Leontovich1961}
M.A.~Leontovich, \emph{{Generalization of the Kramers-Kronig formulas to media with spatial dispersion}}, {\emph{Sov. Phys. JETP} {\bfseries 13} (1961) 634}.

\bibitem{Dolgov:1981}
O.V.~Dolgov, D.A.~Kirzhnits and E.G.~Maksimov, \emph{{On an admissible sign of the static dielectric function of matter}}, \href{https://doi.org/10.1103/RevModPhys.53.81}{\emph{Rev. Mod. Phys.} {\bfseries 53} (1981) 81}.

\bibitem{Streater:1989vi}
R.F.~Streater and A.S.~Wightman, \emph{{PCT, Spin and Statistics, and All That}}, Princeton University Press (1989).

\bibitem{Nussenzveig:1972tcd}
H.M.~Nussenzveig, \emph{{Causality and Dispersion Relations}}, Academic Press, New York (1972).

\bibitem{bogolubov2012general}
N.N.~Bogolubov, A.A.~Logunov, A.I.~Oksak and I.T.~Todorov, \emph{General Principles of Quantum Field Theory}, vol.~10 of \emph{Mathematical Physics and Applied Mathematics}, Springer Science \& Business Media (2012).

\bibitem{Dubovsky:2007ac}
S.~Dubovsky, A.~Nicolis, E.~Trincherini and G.~Villadoro, \emph{{Microcausality in curved space-time}}, \href{https://doi.org/10.1103/PhysRevD.77.084016}{\emph{Phys. Rev. D} {\bfseries 77} (2008) 084016} [\href{https://arxiv.org/abs/0709.1483}{{\ttfamily 0709.1483}}].

\bibitem{Hartman:2015lfa}
T.~Hartman, S.~Jain and S.~Kundu, \emph{{Causality Constraints in Conformal Field Theory}}, \href{https://doi.org/10.1007/JHEP05(2016)099}{\emph{JHEP} {\bfseries 05} (2016) 099} [\href{https://arxiv.org/abs/1509.00014}{{\ttfamily 1509.00014}}].

\bibitem{Hui:2025aja}
L.~Hui, A.~Nicolis, A.~Podo and S.~Zhou, \emph{{Microcausality without Lorentz invariance}}, \href{https://doi.org/10.1007/JHEP07(2025)188}{\emph{JHEP} {\bfseries 07} (2025) 188} [\href{https://arxiv.org/abs/2502.04215}{{\ttfamily 2502.04215}}].

\bibitem{Creminelli:2025rxj}
P.~Creminelli, A.~Longo, B.~Salehian and A.~Zahed, \emph{{Analyticity and positivity of Green{\textquoteright}s functions without Lorentz}}, \href{https://doi.org/10.1007/JHEP06(2026)135}{\emph{JHEP} {\bfseries 06} (2026) 135} [\href{https://arxiv.org/abs/2512.10843}{{\ttfamily 2512.10843}}].

\bibitem{Brillouin1960}
L.~Brillouin, \emph{{Wave Propagation and Group Velocity}}, Academic Press, New York (1960).

\bibitem{Fox:1970cu}
R.~Fox, C.G.~Kuper and S.G.~Lipson, \emph{{Faster-than-light group velocities and causality violation}}, \href{https://doi.org/10.1098/rspa.1970.0093}{\emph{Proc. Roy. Soc. Lond. A} {\bfseries 316} (1970) 515}.

\bibitem{Krotscheck:1978}
E.~Krotscheck and W.~Kundt, \emph{{Causality criteria}}, \href{https://doi.org/10.1007/BF01609447}{\emph{Commun. Math. Phys.} {\bfseries 60} (1978) 171}.

\bibitem{Pu:2009fj}
S.~Pu, T.~Koide and D.H.~Rischke, \emph{{Does stability of relativistic dissipative fluid dynamics imply causality?}}, \href{https://doi.org/10.1103/PhysRevD.81.114039}{\emph{Phys. Rev. D} {\bfseries 81} (2010) 114039} [\href{https://arxiv.org/abs/0907.3906}{{\ttfamily 0907.3906}}].

\bibitem{Gavassino:2023mad}
L.~Gavassino, M.M.~Disconzi and J.~Noronha, \emph{{Dispersion Relations Alone Cannot Guarantee Causality}}, \href{https://doi.org/10.1103/PhysRevLett.132.162301}{\emph{Phys. Rev. Lett.} {\bfseries 132} (2024) 162301} [\href{https://arxiv.org/abs/2307.05987}{{\ttfamily 2307.05987}}].

\bibitem{Hoult:2023clg}
R.E.~Hoult and P.~Kovtun, \emph{{Causality and classical dispersion relations}}, \href{https://doi.org/10.1103/PhysRevD.109.046018}{\emph{Phys. Rev. D} {\bfseries 109} (2024) 046018} [\href{https://arxiv.org/abs/2309.11703}{{\ttfamily 2309.11703}}].

\bibitem{Wang:2023csj}
D.-L.~Wang and S.~Pu, \emph{{Stability and causality criteria in linear mode analysis: Stability means causality}}, \href{https://doi.org/10.1103/PhysRevD.109.L031504}{\emph{Phys. Rev. D} {\bfseries 109} (2024) L031504} [\href{https://arxiv.org/abs/2309.11708}{{\ttfamily 2309.11708}}].

\bibitem{Bros:2001ab}
J.~Bros and H.~Epstein, \emph{{Microcausality and energy-positivity in all frames imply Lorentz invariance of dispersion laws}},  \href{https://arxiv.org/abs/hep-th/0107240}{{\ttfamily hep-th/0107240}}.

\bibitem{Israel:1979wp}
W.~Israel and J.M.~Stewart, \emph{{Transient relativistic thermodynamics and kinetic theory}}, \href{https://doi.org/10.1016/0003-4916(79)90130-1}{\emph{Annals Phys.} {\bfseries 118} (1979) 341}.

\bibitem{Hiscock:1983zz}
W.A.~Hiscock and L.~Lindblom, \emph{{Stability and causality in dissipative relativistic fluids}}, \href{https://doi.org/10.1016/0003-4916(83)90288-9}{\emph{Annals Phys.} {\bfseries 151} (1983) 466}.

\bibitem{Hiscock:1985zz}
W.A.~Hiscock and L.~Lindblom, \emph{{Generic instabilities in first-order dissipative relativistic fluid theories}}, \href{https://doi.org/10.1103/PhysRevD.31.725}{\emph{Phys. Rev. D} {\bfseries 31} (1985) 725}.

\bibitem{Bemfica:2017wps}
F.S.~Bemfica, M.M.~Disconzi and J.~Noronha, \emph{{Causality and existence of solutions of relativistic viscous fluid dynamics with gravity}}, \href{https://doi.org/10.1103/PhysRevD.98.104064}{\emph{Phys. Rev. D} {\bfseries 98} (2018) 104064} [\href{https://arxiv.org/abs/1708.06255}{{\ttfamily 1708.06255}}].

\bibitem{Kovtun:2019hdm}
P.~Kovtun, \emph{{First-order relativistic hydrodynamics is stable}}, \href{https://doi.org/10.1007/JHEP10(2019)034}{\emph{JHEP} {\bfseries 10} (2019) 034} [\href{https://arxiv.org/abs/1907.08191}{{\ttfamily 1907.08191}}].

\bibitem{Bemfica:2019knx}
F.S.~Bemfica, M.M.~Disconzi and J.~Noronha, \emph{{Nonlinear causality of general first-order relativistic viscous hydrodynamics}}, \href{https://doi.org/10.1103/PhysRevD.100.104020}{\emph{Phys. Rev. D} {\bfseries 100} (2019) 104020} [\href{https://arxiv.org/abs/1907.12695}{{\ttfamily 1907.12695}}].

\bibitem{Hoult:2020eho}
R.E.~Hoult and P.~Kovtun, \emph{{Stable and causal relativistic Navier-Stokes equations}}, \href{https://doi.org/10.1007/JHEP06(2020)067}{\emph{JHEP} {\bfseries 06} (2020) 067} [\href{https://arxiv.org/abs/2004.04102}{{\ttfamily 2004.04102}}].

\bibitem{Bemfica:2020zjp}
F.S.~Bemfica, M.M.~Disconzi and J.~Noronha, \emph{{First-order general-relativistic viscous fluid dynamics}}, \href{https://doi.org/10.1103/PhysRevX.12.021044}{\emph{Phys. Rev. X} {\bfseries 12} (2022) 021044} [\href{https://arxiv.org/abs/2009.11388}{{\ttfamily 2009.11388}}].

\bibitem{Gavassino:2021owo}
L.~Gavassino, \emph{{Can We Make Sense of Dissipation without Causality?}}, \href{https://doi.org/10.1103/PhysRevX.12.041001}{\emph{Phys. Rev. X} {\bfseries 12} (2022) 041001} [\href{https://arxiv.org/abs/2111.05254}{{\ttfamily 2111.05254}}].

\bibitem{Gavassino:2023myj}
L.~Gavassino, \emph{{Bounds on transport from hydrodynamic stability}}, \href{https://doi.org/10.1016/j.physletb.2023.137854}{\emph{Phys. Lett. B} {\bfseries 840} (2023) 137854} [\href{https://arxiv.org/abs/2301.06651}{{\ttfamily 2301.06651}}].

\bibitem{Hartman:2017hhp}
T.~Hartman, S.A.~Hartnoll and R.~Mahajan, \emph{{Upper Bound on Diffusivity}}, \href{https://doi.org/10.1103/PhysRevLett.119.141601}{\emph{Phys. Rev. Lett.} {\bfseries 119} (2017) 141601} [\href{https://arxiv.org/abs/1706.00019}{{\ttfamily 1706.00019}}].

\bibitem{Baumann:2015nta}
D.~Baumann, D.~Green, H.~Lee and R.A.~Porto, \emph{{Signs of Analyticity in Single-Field Inflation}}, \href{https://doi.org/10.1103/PhysRevD.93.023523}{\emph{Phys. Rev. D} {\bfseries 93} (2016) 023523} [\href{https://arxiv.org/abs/1502.07304}{{\ttfamily 1502.07304}}].

\bibitem{Grall:2021xxm}
T.~Grall and S.~Melville, \emph{{Positivity bounds without boosts: New constraints on low energy effective field theories from the UV}}, \href{https://doi.org/10.1103/PhysRevD.105.L121301}{\emph{Phys. Rev. D} {\bfseries 105} (2022) L121301} [\href{https://arxiv.org/abs/2102.05683}{{\ttfamily 2102.05683}}].

\bibitem{Creminelli:2022onn}
P.~Creminelli, O.~Janssen and L.~Senatore, \emph{{Positivity bounds on effective field theories with spontaneously broken Lorentz invariance}}, \href{https://doi.org/10.1007/JHEP09(2022)201}{\emph{JHEP} {\bfseries 09} (2022) 201} [\href{https://arxiv.org/abs/2207.14224}{{\ttfamily 2207.14224}}].

\bibitem{Hui:2023pxc}
L.~Hui, I.~Kourkoulou, A.~Nicolis, A.~Podo and S.~Zhou, \emph{{S-matrix positivity without Lorentz invariance: a case study}}, \href{https://doi.org/10.1007/JHEP04(2024)145}{\emph{JHEP} {\bfseries 04} (2024) 145} [\href{https://arxiv.org/abs/2312.08440}{{\ttfamily 2312.08440}}].

\bibitem{Creminelli:2023kze}
P.~Creminelli, M.~Delladio, O.~Janssen, A.~Longo and L.~Senatore, \emph{{Non-analyticity of the S-matrix with spontaneously broken Lorentz invariance}}, \href{https://doi.org/10.1007/JHEP06(2024)201}{\emph{JHEP} {\bfseries 06} (2024) 201} [\href{https://arxiv.org/abs/2312.08441}{{\ttfamily 2312.08441}}].

\bibitem{Creminelli:2024lhd}
P.~Creminelli, O.~Janssen, B.~Salehian and L.~Senatore, \emph{{Positivity bounds on electromagnetic properties of media}}, \href{https://doi.org/10.1007/JHEP08(2024)066}{\emph{JHEP} {\bfseries 08} (2024) 066} [\href{https://arxiv.org/abs/2405.09614}{{\ttfamily 2405.09614}}].

\bibitem{Cheung:2026wrz}
T.~Cheung, C.~Duaso-Pueyo, D.~Stefanyszyn and D.-G.~Wang, \emph{{General Spin-1 Dynamics During Inflation: Microcausality Constraints and Cosmological Collider Signals}},  \href{https://arxiv.org/abs/2610.10693}{{\ttfamily 2610.10693}}.

\bibitem{Heller:2022ejw}
M.P.~Heller, A.~Serantes, M.~Spali{\'n}ski and B.~Withers, \emph{{Rigorous Bounds on Transport from Causality}}, \href{https://doi.org/10.1103/PhysRevLett.130.261601}{\emph{Phys. Rev. Lett.} {\bfseries 130} (2023) 261601} [\href{https://arxiv.org/abs/2212.07434}{{\ttfamily 2212.07434}}].

\bibitem{Heller:2023hydrohedron}
M.P.~Heller, A.~Serantes, M.~Spali{\'n}ski and B.~Withers, \emph{{The space of transport coefficients allowed by causality}}, \href{https://doi.org/10.1038/s41567-024-02635-5}{\emph{Nature Phys.} {\bfseries 20} (2024) 1948} [\href{https://arxiv.org/abs/2305.07703}{{\ttfamily 2305.07703}}].

\bibitem{Rogosinski1932}
W.~Rogosinski, \emph{{{\"U}ber positive harmonische Entwicklungen und typisch-reelle Potenzreihen}}, \href{https://doi.org/10.1007/BF01186552}{\emph{Math. Z.} {\bfseries 35} (1932) 93}.

\bibitem{Robertson1935}
M.S.~Robertson, \emph{{On the coefficients of a typically-real function}}, \href{https://doi.org/10.1090/S0002-9904-1935-06147-6}{\emph{Bull. Amer. Math. Soc.} {\bfseries 41} (1935) 565}.

\bibitem{Creminelli:2026dispersion}
P.~Creminelli, M.~Mancini and W.P.~McBlain, \emph{{Microcausality Bounds on Dispersion Laws}},  \href{https://arxiv.org/abs/2611.17407}{{\ttfamily 2611.17407}}.

\bibitem{KreinNudelman:1977}
M.~Kre{\u{\i}}n and A.~Nudel'man, \emph{{The Markov Moment Problem and Extremal Problems}}, American Mathematical Society (1977), \href{https://doi.org/10.1090/mmono/050}{10.1090/mmono/050}.

\bibitem{Son:2002zn}
D.T.~Son, \emph{{Low-energy quantum effective action for relativistic superfluids}},  \href{https://arxiv.org/abs/hep-ph/0204199}{{\ttfamily hep-ph/0204199}}.

\bibitem{Nicolis:2011pv}
A.~Nicolis and F.~Piazza, \emph{{Spontaneous Symmetry Probing}}, \href{https://doi.org/10.1007/JHEP06(2012)025}{\emph{JHEP} {\bfseries 06} (2012) 025} [\href{https://arxiv.org/abs/1112.5174}{{\ttfamily 1112.5174}}].

\bibitem{Heller:2020hnq}
M.P.~Heller, A.~Serantes, M.~Spali{\'n}ski, V.~Svensson and B.~Withers, \emph{{Convergence of hydrodynamic modes: insights from kinetic theory and holography}}, \href{https://doi.org/10.21468/SciPostPhys.10.6.123}{\emph{SciPost Phys.} {\bfseries 10} (2021) 123} [\href{https://arxiv.org/abs/2012.15393}{{\ttfamily 2012.15393}}].

\bibitem{Gavassino:2026acausal}
L.~Gavassino, \emph{{How acausal equations emerge from causal dynamics}},  \href{https://arxiv.org/abs/2604.07031}{{\ttfamily 2604.07031}}.

\bibitem{Brants:2026causalUV}
R.~Brants, \emph{{Causal UV completions of relativistic hydrodynamics}},  \href{https://arxiv.org/abs/2605.21377}{{\ttfamily 2605.21377}}.

\bibitem{Vladimirov1966}
V.S.~Vladimirov, \emph{{Methods of the Theory of Functions of Many Complex Variables}}, MIT Press, Cambridge, MA (1966).

\bibitem{Grozdanov:2019uhi}
S.~Grozdanov, P.K.~Kovtun, A.O.~Starinets and P.~Tadi{\'c}, \emph{{The complex life of hydrodynamic modes}}, \href{https://doi.org/10.1007/JHEP11(2019)097}{\emph{JHEP} {\bfseries 11} (2019) 097} [\href{https://arxiv.org/abs/1904.12862}{{\ttfamily 1904.12862}}].

\bibitem{Esposito:2026diu}
A.~Esposito, A.~Nicolis and R.~Penco, \emph{{Effective field theories for material media}}, \href{https://doi.org/10.1016/j.physrep.2026.11.001}{\emph{Phys. Rept.} {\bfseries 1222} (2027) 1} [\href{https://arxiv.org/abs/2607.06666}{{\ttfamily 2607.06666}}].

\bibitem{Landau:1984ecm}
L.D.~Landau, E.M.~Lifshitz and L.P.~Pitaevskii, \emph{{Electrodynamics of Continuous Media}}, Pergamon Press, Oxford, 2nd~ed. (1984).

\bibitem{AkhiezerKrein1962}
N.I.~Akhiezer and M.G.~Kre{\u\i}n, \emph{{Some Questions in the Theory of Moments}}, vol.~2 of \emph{Translations of Mathematical Monographs}, American Mathematical Society, Providence, RI (1962).

\bibitem{diaconis2004markov1}
P.~Diaconis and D.~Freedman, \emph{{The Markov moment problem and de Finetti's theorem: Part I}}, \href{https://doi.org/10.1007/s00209-003-0636-6}{\emph{Math. Z.} {\bfseries 247} (2004) 183}.

\bibitem{Akhiezer1965}
N.I.~Akhiezer, \emph{{The Classical Moment Problem and Some Related Questions in Analysis}}, Oliver \& Boyd, Edinburgh (1965).

\bibitem{schmudgen2017moment}
K.~Schm{\"u}dgen, \emph{The Moment Problem}, vol.~277 of \emph{Graduate Texts in Mathematics}, Springer, Cham, Switzerland (2017), \href{https://doi.org/10.1007/978-3-319-64546-9}{10.1007/978-3-319-64546-9}.

\bibitem{Bellazzini:2020cot}
B.~Bellazzini, J.~Elias~Mir{\'o}, R.~Rattazzi, M.~Riembau and F.~Riva, \emph{{Positive moments for scattering amplitudes}}, \href{https://doi.org/10.1103/PhysRevD.104.036006}{\emph{Phys. Rev. D} {\bfseries 104} (2021) 036006} [\href{https://arxiv.org/abs/2011.00037}{{\ttfamily 2011.00037}}].

\bibitem{Apostolidis:2026mvb}
T.~Apostolidis, A.~Longo, B.~Salehian and L.~Santoni, \emph{{Toward the CompactHedron: Bounds on Neutron Stars from Microscopic Principles}},  \href{https://arxiv.org/abs/2609.36025}{{\ttfamily 2609.36025}}.

\bibitem{Weisstein:Sine}
E.W.~Weisstein, ``{Sine}.'' Wolfram MathWorld, \url{https://mathworld.wolfram.com/Sine.html}.

\bibitem{Luenberger1969}
D.G.~Luenberger, \emph{{Optimization by Vector Space Methods}}, John Wiley \& Sons, New York (1969).

\bibitem{BoydVandenberghe2004}
S.~Boyd and L.~Vandenberghe, \emph{{Convex Optimization}}, Cambridge University Press (2004).

\bibitem{Caron-Huot:2020cmc}
S.~Caron-Huot and V.~Van~Duong, \emph{{Extremal Effective Field Theories}}, \href{https://doi.org/10.1007/JHEP05(2021)280}{\emph{JHEP} {\bfseries 05} (2021) 280} [\href{https://arxiv.org/abs/2011.02957}{{\ttfamily 2011.02957}}].

\bibitem{Withers:2018srf}
B.~Withers, \emph{{Short-lived modes from hydrodynamic dispersion relations}}, \href{https://doi.org/10.1007/JHEP06(2018)059}{\emph{JHEP} {\bfseries 06} (2018) 059} [\href{https://arxiv.org/abs/1803.08058}{{\ttfamily 1803.08058}}].

\bibitem{Grozdanov:2019kge}
S.~Grozdanov, P.K.~Kovtun, A.O.~Starinets and P.~Tadi{\'c}, \emph{{Convergence of the Gradient Expansion in Hydrodynamics}}, \href{https://doi.org/10.1103/PhysRevLett.122.251601}{\emph{Phys. Rev. Lett.} {\bfseries 122} (2019) 251601} [\href{https://arxiv.org/abs/1904.01018}{{\ttfamily 1904.01018}}].

\end{thebibliography}\endgroup

\end{document}